\documentclass[useAMS,usenatbib]{mn2e}
\usepackage{epsfig}
\newcommand\msun{\rm{M_{\odot}}}
\title[Titolo]{Hydrodynamical simulations of Galactic fountains II: evolution of multiple fountains}
\author[C. Melioli, F. Brighenti, A. D'Ercole, E.M. de Gouveia Dal Pino]
{C. Melioli,$^{1}$\thanks{E-mail:
claudio.melioli@oabo.inaf.it; fabrizio.brighenti@unibo.it; annibale.dercole@oabo.inaf.it;
dalpino@astro.iag.usp.br}
F. Brighenti$^{2}$, A.D'Ercole$^{1}$ and E.M. de Gouveia Dal Pino$^{3}$\\
$^{1}$INAF-Osservatorio Astronomico di Bologna, via Ranzani 1, 40126 Bologna, Italy\\
$^{2}$Dipartimento di Astronomia, Universit\`a di Bologna, via Ranzani 1, 40126 Bologna, Italy\\
$^{3}$IAG, Universidade de S\~ao Paulo, Rua do Mat\~ao 1226, 05508-090 S\~ao Paulo, Brazil}
\begin{document}

\date{Accepted ... Received ...; in original form ...}

\pagerange{\pageref{firstpage}--\pageref{lastpage}} \pubyear{2002}

\maketitle

\label{firstpage}

\begin{abstract}
  The ejection of gas out of the disk in late-type galaxies is related
  to star formation and is mainly due to the explosion of Type II
  supernovae. In a previous paper, we considered the evolution of a
  single Galactic fountain, that is, a fountain powered by a single
  supernova cluster. Using 3D hydrodynamical simulations, we studied
  in detail the fountain flow and its dependence with several factors,
  such as the Galactic rotation, the distance to the Galactic center,
  and the presence of a hot gaseous halo. As a natural follow-up, the
  present paper investigates the dynamical evolution of multiple
  generations of fountains generated by $\sim 100$ OB associations.
  We have considered the observed size-frequency distribution of young
  stellar clusters within the Galaxy in order to appropriately fuel
  the multiple fountains in our simulations. Most of the results of
  the previous paper have been confirmed, like for example the
  formation of intermediate velocity clouds above the disk by the
  multiple fountains. Also, the present work confirms the localized
  nature of the fountain flows: the freshly ejected metals tend to
  fall back close to the same Galactocentric region where they are
  delivered. Therefore, the fountains do not change significantly the
  radial profile of the disk chemical abundance.  The multiple
  fountains simulations also allowed to consistently calculate the
  feedback of the star formation on the halo gas. We found that the
  hot gas gains about 10 \% of all the SNII energy produced in the
  disk.  Thus, the SN feedback more than compensate for the halo
  radiative losses and allow for a quasi steady-state disk-halo
  circulation to exist.  Finally, we have also considered the
  possibility of mass infall from the intergalactic medium and its
  interaction with the clouds that are formed by the fountains.
  Though our simulations are not suitable to reproduce the slow
  rotational pattern that is typically observed in the halos around
  the disk galaxies, they indicate that the presence of an external
  gas infall may help to slow down the rotation of the gas in the
  clouds and thus the amount of angular momentum that they transfer to
  the coronal gas, as previously suggested in the literature.

\end{abstract}


\section{Introduction}
 \label{sec:introduction}

Galactic fountains (GFs) are believed to occur in
the Milky Way as well as in the other disk galaxies. A GF takes
place when the Type II supernovae (SNe II) belonging to an OB
association are sufficiently numerous to create a superbubble and to
drive its expansion above the scale height of the gaseous disk
\citep{shafi76, breg80, kahn81}. As the break out occurs
\citep{mamcno89,kmck92}, the gas heated by the SNe II falls back
to the Galactic plane as cold gas, mostly concentrated in clouds
formed through thermal instabilities.

The GF mechanism is thought to be linked to a number of issues which
have been discussed in some detail in \citet [][thereafter Paper
I]{mel08}.  Here we briefly recall them: $i$) the formation of the
observed thick H {\sevensize I} layer having a mass of 1/10 of the
total H {\sevensize I}, and rotating with velocity about 20-50 km
s$^{-1}$ smaller than that of the gas on the plane; $ii$) the presence
of high velocity clouds (HVCs) and/or intermediate velocity clouds
(IVCs). It is important to understand whether these clouds are formed
by GFs, or represent external gas accreting on the disk
\citep{safra08}; $iii$) the presence of giant holes in the H
{\sevensize I} disk of many galaxies, which can be produced by
associations of O and B stars exploding in a correlated fashion and
giving rise to superbubbles; $iv$) the chemical evolution of the
Galaxy, as in principle GFs might be able to move SN II ejecta
relatively far from the place where they have been produced, affecting
the gradients of $\alpha$-elements; $v$) the diffuse, soft ($T \la$ 1
keV), extraplanar X-ray emission detected in some spiral galaxies
which is due to hot gas extending up to 10 kpc above the galactic
disk. This gas is most likely blown out by star forming regions in the
disk, as the X-ray luminosity correlates with the star formation rate
and the hot gas is oxygen rich, a signature of SN II enrichment.

Given the richness of the phenomena involved in the GFs, a number of
papers has been devoted to this argument.  A detailed analytical
description of the motion of a fluid element in a GF from the rotating
disk to the halo and back, predicting the location where it would
return on the disk was first presented by \citet{kahn91} and
\citet{kahn93}.  \citet{deav00} and \citet{deav01} performed several
3D hydrodynamical simulations of the gas in the Milky Way in order to
account for the collective effects of supernovae on the structure of
the interstellar medium (ISM). In particular, \citet{deav00} discussed
in detail for the first time the fraction of material that results in
the formation of IVCs. More recently, further 3D MHD simulations were
carried out by \citet{deav05}. They found that gas transport into the
halo as well as the mean volume filling factor of the hot phase in the
disk is not significantly affected by the presence of a initial disk
parallel magnetic field \citep[see also][]{tom98}.  In order to reach a
high spatial resolution, all these simulations consider only small
disk regions (typically 1 kpc$^2 \times$ 10 kpc) and assume plane
parallel stratified ISM.  \citet{kor99} have included in their 3D
numerical simulations the effects of the differential rotation and the
magnetic field of the galactic disk, but again considered a small
computational domain (500 pc$^2$ $\times$ 1 kpc). In all the simulations
above any dependence of the variables with the Galactocentric distance
is neglected. The lack of an extended grid makes the entire
development of a disk-halo-disk cycle very difficult to follow because
part of the mass is lost through the grid boundaries.

The simulations by \citet{frbi06} followed a different strategy.
They considered the whole Galaxy and calculated the orbits of gas
clouds assumed to be like bullets ejected (by SNe II) from the
disk that move ballistically up into the halo and then return to the
disk. However, hydrodynamical effects have been essentially
neglected in these models.

In order to study the large scale motion of the ISM and the halo gas
and follow the complete dynamical and thermal evolution of the gas
lifted by the galactic fountains, the whole, differentially rotating
Galaxy must be considered.  In Paper I we have presented 3D
hydrodynamical simulations of the evolution of a single GF (SGF),
i.e. a fountain generated by only one OB association with 100 SNe.
There, we investigated the influence on the SGF dynamics of several
factors, such as the Galactocentric distance, the Galactic rotation,
and the presence of a gaseous halo and/or a disk neutral hydrogen
layer.  We have found that a SGF may eject material up to $\sim 2$ kpc
before it collapses back in the disk, mostly in form of dense, cold
clouds and filaments, with velocities which are compatible with those
of the IVCs. Contrary to the common expectation, the gas lifted up by
the SGF tends to move toward the centre rather than toward the
outskirts of the disk because part of its angular momentum is
transferred to the halo and/or the upper layers of the disk. However,
most of the GF gas falls back on the disk within a radial distance
$\Delta R\sim 0.5$ kpc from the place where the fountain
originated. This localized circulation of disk gas agrees with recent
chemical models of the Milky Way which assume that the SNII metals
enrich only the local ISM. \citep{cemafr07}.

The present paper is devoted to multiple Galactic fountains (MGFs) -
i.e. fountains generated by multiple OB associations - in which the gas
flow is due to the continuous star formation in the Galactic
disk. Thus, this paper represents a natural extension
of the analysis of Paper I on single fountains.  To keep
the computational costs under reasonable limits, we have followed in
detail only a limited volume of the Galaxy where the MGF occurs; the
remaining Galactic volume was mapped at a lower resolution. Although a
very accurate description of the different ISM phases \citep[as
in][]{deav00,deav01} is hampered by the limited resolution resulting
from our approach, the global evolution of the fountains can be
satisfactorily calculated. We neglected the effect of magnetic field;
as shown by the aforementioned papers by \citet{kahn91,kahn93,deav05}
it does not strongly affect the global fountain dynamics.

An important motivation to run simulations of MGFs is to investigate
the mass and energy exchange between the star forming disk and the hot
gaseous halo surrounding the Galaxy. The efficiency of the SN feedback
is a key, yet poorly understood issue in theories of galactic star
formation, as we will outline later \citep[c.f.][]{nav97}. We find
that a non-negligible fraction of the SN energy is indeed injected
in the halo (Section 3.4).

A further reason to consider multiple fountains is to explore
the interaction with external gas infalling
onto the disk.  Indeed, such  gas is needed in current galaxy
evolution models in order to maintain star formation rates similar
to those observed in normal spiral galaxies \citep
[e.g.][]{takhir00,semcom02}, as well as to explain the observed
patterns in the chemical evolution of the Milky Way
\citep{chi02,gei02}. \citet{frbi07} detected inflowing HI clouds in
the nearby spiral galaxies NGC 891 and NGC 2403. They argue that
these clouds are a clue to a substantial and otherwise mostly hidden
halo gas accretion interacting with GFs. While the required
accretion rate is comparable to the star formation rate, the
infalling gas must have low  angular momentum \citep{frbi06}.

In the following sections, we  outline the basic characteristics and
the setup of our numerical model (Sec. 2), then  examine a reference
model for the investigation of multiple fountains (MGFs) formation in
the Galaxy at a distance of 8.4 kpc from the galactic centre (Sec.
3) and compare these results with a model where the MGFs occur at $R=4.5$ kpc
(Sec. 4). Finally, in Sec. 5, we consider the interaction of MGFs
with infalling gas from the intergalactic medium and in Sec. 6, we
draw our conclusions.

\section{the model}
 \label{sec:models}
 The model of the Milky Way, as well as the assumptions relative to
 the SN II explosions belonging to single associations, are fully
 described in Paper I. Here we briefly recapitulate them and
 discuss in detail how we take into account the stellar explosions
 occurring in different OB associations.
\subsection{The Galaxy model}
 \label{subsec:galmod}

 The ISM in our model is made up of three components - namely
 molecular (H$_2$), neutral (H {\sevensize I}) and ionized (H
 {\sevensize II}) hydrogen - all following the spatial distribution
 given by \citet{wolf03} (see also Paper I).  The ISM is initially set
 in rotational equilibrium in the Galactic gravitational potential due
 to the summation of the dark matter halo, the bulge and the disk
 contributions, as described by \citet{nfw96}, \citet{her90}, and by a
 flattened King profile \citep{brma96}, respectively. A hot isothermal
 gas halo in hydrostatic equilibrium with the Galactic potential well
 is also added.  This halo has a central density $\rho_{\rm
   0,h}=2.17\times 10^{-27}$ g cm$^{-3}$ and a temperature $T_{\rm
   h}=7\times 10^6$ K.

\subsection{Supernovae explosions}
 \label{subsec:snexp}
\subsubsection{Single associations}
\label{subsubsec:sngas}

It is known that the great majority of the massive stars form in
clusters \citep[e.g.][and references therein]{watho02}. In Paper I,
we defined a ``single fountain'' as a fountain powered by the SNe II
present in a single cluster\footnote{For the sake of simplicity, by
SNe II we mean here all the core collapse explosions.}.  The OB
association driving the fountain was assumed to be totally housed
inside a single zone of the highest resolution level of our adaptive
grid (see section \ref{subsec:nummet}).  The SNe II do not explode
all at the same time because of the different mass of their
progenitors. Thus the energy and the mass delivered by all the
explosions are released over a time interval of $\Delta t$=30 Myr
(the lifetime of an 8 $M_{\odot}$ star, which is the least massive
SN II progenitor). The stellar explosions inject mass and energy at
the rates $\dot M ={\cal R}M_{\rm ej}$ and $L_{\rm w}={\cal R}E_0$,
respectively. Here $M_{\rm ej}=16$ $M_{\odot}$ and $E_0=10^{51}$ erg
are the mean mass and energy released by a single explosion,
respectively; ${\cal R}=N_{\rm SN}/\Delta t$ is the SN II rate, and
$N_{\rm SN}$ is the total number of SN II explosions occurring in
the association.  $L_{\rm w}$ represents the mechanical luminosity
of the wind driven by the SNe II and inflating a superbubble. If the
luminosity is larger than a critical value $L_{\rm cr}$, the
superbubble breaks out of the disk forming a chimney in the ISM and
gives rise to the GF (cf. Paper I and references therein). In section
\ref{subsec:nummet} we describe as $L_{\rm w}$ is implemented
into the numerical grid.

\subsubsection{Multiple associations}
\label{subsubsec:mltas}

At the observed rate (see below), several SGFs may occur
sufficiently close to each other in time and space to mutually
interact.  This interaction obviously modifies the general dynamics of
the SN II ejecta of each single fountain.  Following with an adequate
spatial resolution all the SN II explosions occurring in the whole
disk is beyond our current computational capabilities.  Thus,
throughout the simulations of multiple fountains, we focus in
particular on a circular area (hereafter referred to as the ``active
area'') of the Galactic disk of 8 kpc$^2$ which is resolved up to the
maximum level of grid refinement (this finest grid extends vertically
up to $z=3.2$ kpc).  We allow the SNe II to occur only in this area
over a period $T=200$ Myr. It must be emphasized that the star
formation rate (and the associated SN rate) is not computed
self-consistently with the gas evolution. Instead, we assume a mean
SN rate adequately scaled from the rate $1.4\times 10^{-2}$ yr$^{-1}$
relative to the whole Galaxy \citep{cap97}. It turns out that during
the time $T$ of our simulations $N_{\rm tot}=4.2\times 10^4$ SNe II
are expected to explode in the active area of the Galactic disk (see,
however, the end of this section).

Let us assume $f(N_{\rm SN})dN_{\rm SN}$ to be the number of stellar
clusters with a number of supernova progenitors between $N_{\rm SN}$
and $N_{\rm SN}+dN_{\rm SN}$. $f(N_{\rm SN})$ follows a power law
distribution \citep{hili05},

\begin{equation}
f(N_{\rm SN})\propto N_{\rm SN}^{-2},\quad\qquad N_{\rm min }<N_{\rm
SN}<N_{\rm max}.
\end{equation}
We normalize $f(N_{\rm SN})$ in order to get
\begin{equation}
\label{eq:ntot}
\int_{N_{\rm min}}^{N_{\rm max}}N_{\rm SN}f(N_{\rm SN})dN_{\rm SN}=N_{\rm tot}.
\end{equation}
\noindent
This normalization can be done once $N_{\rm min}$ and $N_{\rm max}$ have been
chosen (see below in this section).

We choose bins in $N_{\rm SN}$ such  that

\begin{equation}
\int_{N^i_{\rm SN}}^{N^{i+1}_{\rm SN}}f(N_{\rm SN})dN_{\rm SN}=1;
\end{equation}
\noindent
we then assume that such a cluster contains $0.5(N^i_{\rm
SN}+N^{i+1}_{\rm SN})$ SNe II.  Clearly, for $i=1$ we have $N^1_{\rm
SN}=N_{\rm min}$.

\begin{figure}
\begin{center}
\psfig{figure=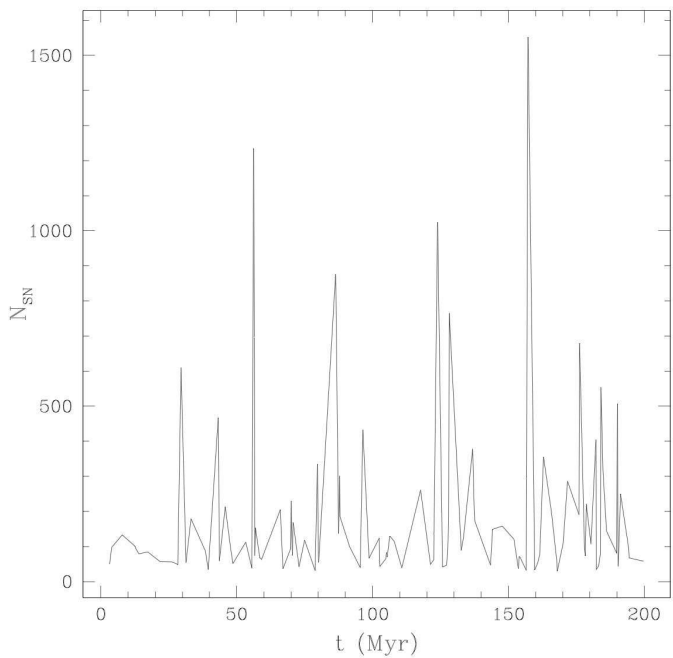,width=0.5\textwidth}
\end{center}
\caption{ Temporal histogram of the sequence of stellar clusters
fuelling our models of multiple Galactic fountains. The y-axis
reports the number of SNe II of each cluster; the x-axis shows the
time in Myr. } \label{fig:clu}
\end{figure}

Once obtained the number and the richness of the clusters, we place
them within the active area. This is made associating randomly to
each $i$-th cluster a position $P_j$ within the active area and a
time $t_j$ in the range $0<t_j<T$. As the time proceeds, the
supernovae belonging to the $i$-th cluster start to explode at $P_j$
at the time $t_j$ (and continue to explode for 30 Myr, as outlined
in section \ref{subsubsec:sngas}). The sequel of the SN II
``bursts'' is shown in Fig. \ref{fig:clu}. The energy provided by
the $i$-th burst is $E_0N^i_{\rm SN}$.

To choose $N_{\rm min }$ and $N_{\rm max}$, we simply assume $N_{\rm
  max}$=2000 from \citet{hili05}. With regard to $N_{\rm min }$, although in
principle it can be as low as $N_{\rm min }=1$, we assume $N_{\rm
min }=30$.  At a distance of $R=8.5$ kpc, which is the
Galactocentric distance of the centre of the active area in our
reference model (cf. section \ref{sec:mrefmod}), GFs can form only
when $L_{\rm w}\ga L_{\rm cr}= 3\times 10^{37}$ erg s$^{-1}$ (cf.
Paper I), that is for clusters richer than $N_{\rm SN}\ga 30$.  In
order to capture the physics of interest here, we have thus adopted
 $N_{\rm min}=30$.  As a $caveat$, we stress that with this
choice of $N_{\rm min}$ the actual number of SN II explosions
occurring after $T=200$ Myr is nearly $N_{\rm tot}/2$, i.e.,  half
of those expected. Although this is not important from a dynamical
point of view, it influences the amount of metals set in circulation
by the fountains. We have to remember this when discussing the
chemical pollution of the disk.

\subsection{Numerical Setup}
\label{subsec:nummet}

We use a modified version of the numerical adaptive mesh refinement
(AMR) YGUAZUa hydrodynamical code \citep{raga00,raga02}.
The radiative losses are
calculated assuming solar abundances and by using 
non-equilibrium ionization species abundances (for HI, HII, CII, CIII,
CIV, OI, OII and OIII) when $T\ge 10^6$ K. For higher temperatures,
where the collisional ioniziation equilibrium is often a good approximation,
we use an equilibrium cooling function taken
from \citet{sudo93}. More details
can be found in Paper I.
We enforce the maximum grid resolution only within the active volume
defined in the previous section (given the symmetry of the problem,
our computational volume encompasses only the ``upper'' half of the
Galaxy). As a reasonable compromise between the need of a large
active surface and a good spatial resolution within it, we chose an
area of 8 kpc$^2$ and a minimum mesh size of 50 pc. The other three
coarser grid levels have mesh sizes of 100 pc, 200 pc and 400 pc,
respectively.

\begin{figure*}
\begin{center}
\psfig{figure=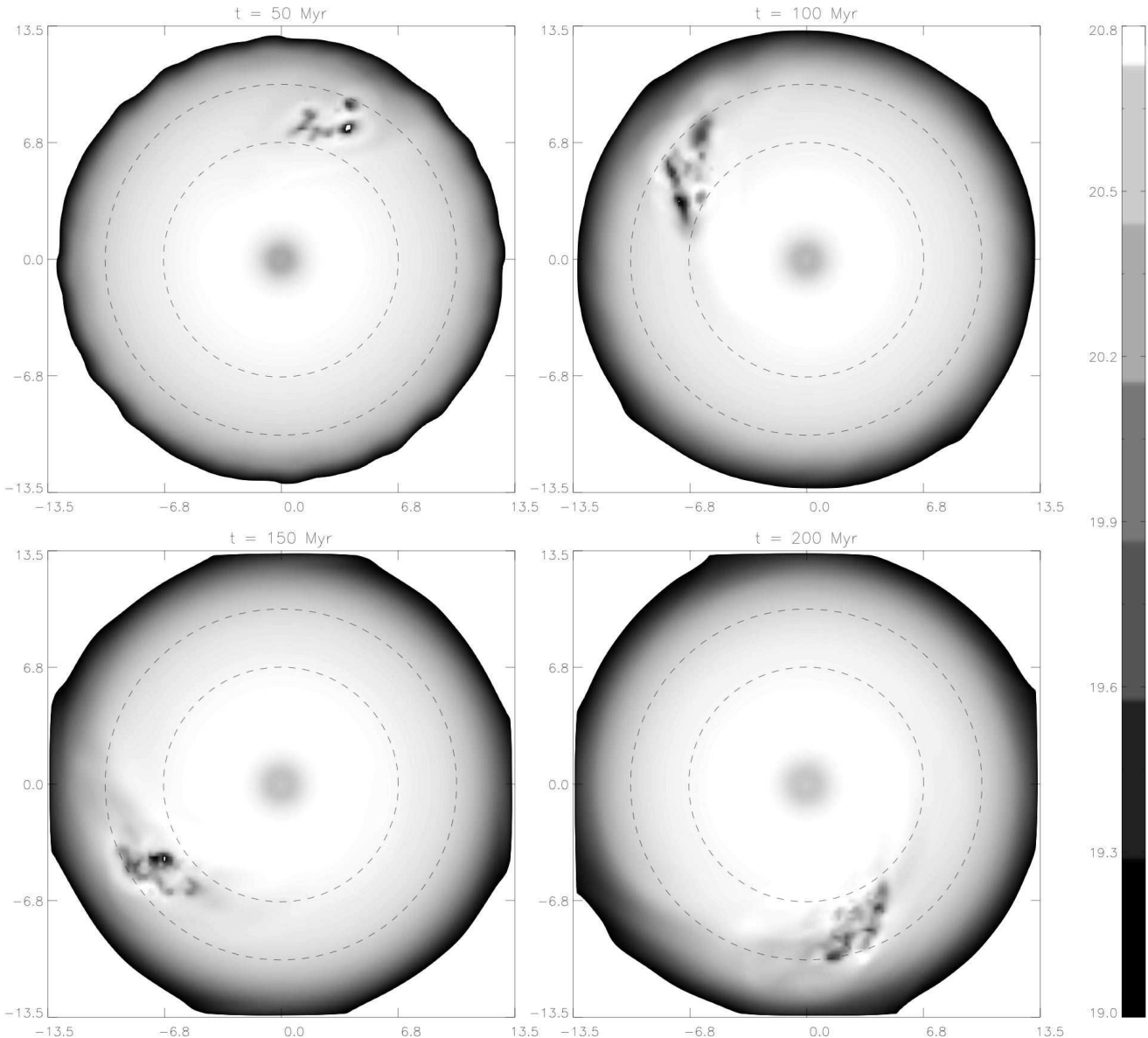,width=1.0\textwidth}
\end{center}
\caption{Face-on view of the active area for the reference model
  (RM). Column density along the $z$ direction of the ISM of the
  Galactic disk at several times. The two superimposed dashed circles
  have radii of 6.8 kpc and 10 kpc and they delimit the ring within
  which the active disk area moves.  The logarithmic column density
  scale is given in cm$^{-2}$ and the spatial scale along the $x$ and $y$ directions is in kpc.}
\label{fig:coldenism}
\end{figure*}

The rate of mass and energy injection by SNe II belonging to every single
association is described in section
\ref{subsubsec:sngas}.  Because the numerical grid is at rest in space while
the Galaxy rotates, each OB association occurring in the active
area moves along a circle on the $z=0$ plane.
At each time we place the 
energy and mass sources in the numerical
cell of the finest grid transited by the association at that time.
We stress that while we assume that all the OB association are located at $z=0$,
SNII progenitors are observed to have a vertical distribution with scale height
of $\sim 100$ pc. As our finest grid has a size of 50 pc, in order to investigate 
the effect of the vertical position on the fountain evolution we must locate OB associations 
only two zones above the equatorial plane. Such a small displacement is not expected to
generate appreciable differences in the calculated flow evolution.

\section{The reference model}
 \label{sec:mrefmod}

 In this section we discuss the reference model (RM), in which the active
 area is centred at the Galactocentric distance $R=8.4$ kpc.

\subsection{ISM in the disk}
\label{subsec:ism}

Figure \ref{fig:coldenism} shows the time evolution of the gas column
density of the ISM of the disk along the direction perpendicular to
the Galactic plane (the $z$-direction). The two dashed circles in the
panels delimit the region within which the active area evolves as the
Galaxy rotates.  There are several cavities whose lifetime is given by
the duration of the SN activity in each association ($\sim 30$ Myr)
added to the time needed for the replenishment to occur at the end of
such activity. The replenishment time is approximately given by
$t_{\rm rep}\sim r_{\rm c}/c$, where $r_{\rm c}$ is the maximum radius
attained by the cavities, and $c$ is the sound speed of the ambient
medium.  It turns out that $r_{\rm c}\sim 0.7$ kpc for the largest
cavities, while $c\sim 2\times 10^6$ cm s$^{-1}$, and thus $t_{\rm
  rep}\sim 37$ Myr. In conclusion, the lifetime of the holes into the
disk ISM is $\la 70$ Myr.  This lifetime and the occurrence rate of
the OB associations both conspire to avoid a substantial overlapping
of the holes and the area of interest is never totally deprived of its
ISM. Given the differential rotation of the Galaxy, the holes become
elongated with positive pitch angles (angle between the major axis of
the hole and the tangent to a circle at that galactic radius), as
described by the models of \citet{palo90}. As a further effect of the
disk differential rotation, the holes created at earlier times lag
behind the recent ones, tending to form a sort of queue. This latter
effect, however, arises because our active region is limited. Had the active
area covered the whole disk, its appearance would be similar to NGC
2403 \citep{frat02} and other typical spirals.

In a recent paper \citet{boom08} presented a detailed study of the H
{\sevensize I} holes and HVCs in the spiral galaxy NGC 69456. It is
interesting to compare our results to their findings. \citet{boom08}
evaluate an average radius of the holes of 0.6 kpc and a maximum
lifetime of about 80 Myr; both these figures are very close to the
values obtained in our simulation. Those authors also find that the
hole covering factor drops sharply toward the smallest galactic radii,
an effect ascribed to the drop of the H {\sevensize I} column density
as well as to the stronger shear which shortens the hole
lifetimes. Smaller dimensions and shorter lifetimes at smaller radii
are also met in our models, and we find that the reason is given by
the larger pressure of the ambient gas which contrasts the bubble
expansion and favours the hole replenishment
(cf. sect. \ref{sec:mr45mod}).  Finally, \citet{boom08} estimate an
upper limit of about $10^7$ $\msun$ for the average H {\sevensize I}
mass missing from each hole. In our simulation we have 3-5 holes
present at every time (cf. Fig. \ref{fig:coldenism}), while the disk
gas lifted at any time is a few $10^7$ $\msun$
(cf. sect. \ref{subsec:feedback}), thus giving roughly $5\times 10^6$
$\msun$ per hole, consistently with the upper limit estimated by
\citet{boom08}. In conclusion, despite our one-fluid approximation of
the multi-phase ISM and the galactic model tailored on the Milky Way
rather than on NGC 69456, our results are in encouraging agreement
with the observational data.

\begin{figure*}
\begin{center}
\psfig{figure=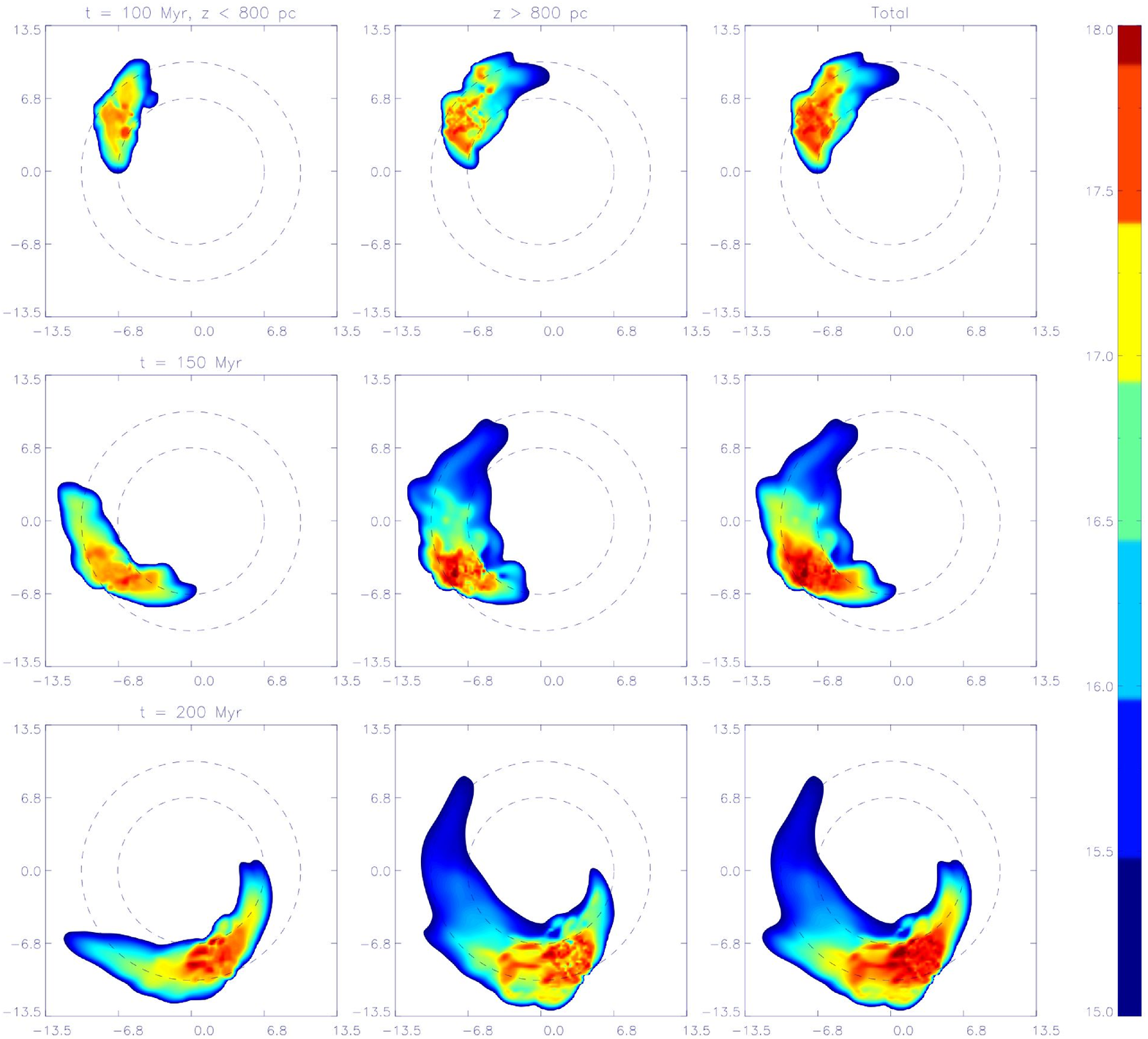,width=1.0\textwidth}
\end{center}
\caption{ Column density along the $z$ direction of the SN II ejecta
  for the reference model.  Each row corresponds to a different
  time. The left and central panels correspond to the column densities
  calculated in the intervals $z<0.8$ kpc and $z>0.8$ kpc,
  respectively, where $z=0.8$ kpc corresponds to the location of the
  disk/halo transition. The right column shows the total column
  densities.  The two circles have the same meaning as in
  Fig. \ref{fig:coldenism}. The logarithmic column density scale is
  given in cm$^{-2}$ and the spatial scale along the $x$ and $y$ directions is in kpc.  }
\label{fig:colden}
\end{figure*}

\begin{figure*}
\begin{center}
\psfig{figure=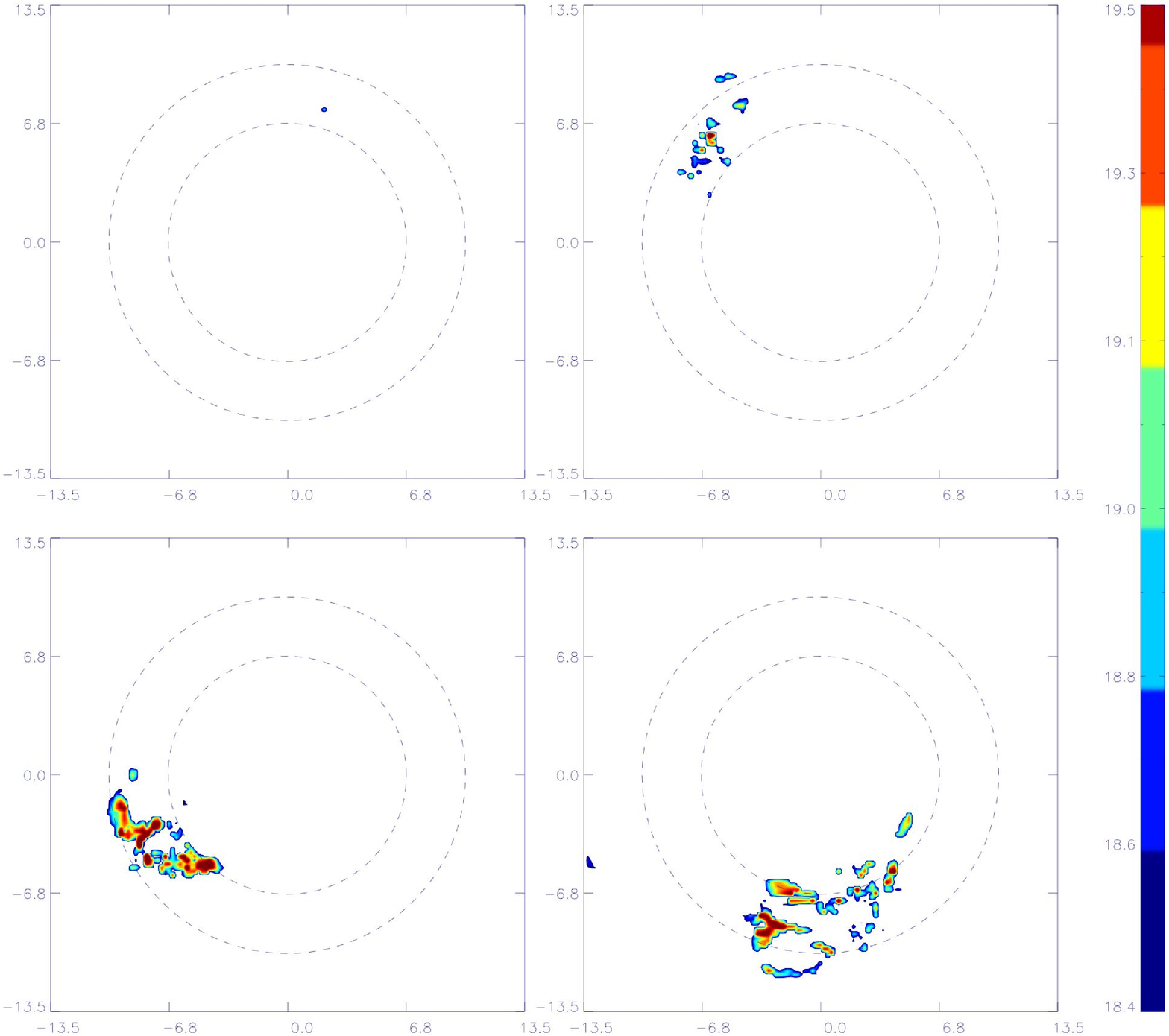,width=1.0\textwidth}
\end{center}
\caption{ Column density along the $z$ direction of the gas falling
  back to the disk, for the reference model.  The times are the same
  as in Fig. \ref{fig:coldenism}, as well as the meaning of the two
  circles.  As apparent from the figure, we consider only gas
  condensed into clouds and filaments. It is also clear that the
  clouds remain essentially within the circular sector in which the
  active area is located. The logarithmic column density scale is
  given in cm$^{-2}$ and the spatial scale along the $x$ and $y$ directions is in kpc.  }
\label{fig:vneg}
\end{figure*}

\subsection{Gas circulation}
\label{subsec:circulation}

Figure \ref{fig:colden} represents a sort of tomography at different
times of the column density of the SN II ejecta along the $z$
direction.  The total column densities are shown in the right
panels; the left and right panels illustrate the contributions of
ejecta located below and above $z=800$ pc (the height at which the
disk/halo transition is located for a radial distance of $R=8.4$
kpc), respectively. The dense spots present in the first column of
the panels are due to recent SN explosions (see, for example, the
top-left panel). Instead, the more structured distribution visible
in the middle column is due to ejecta trapped and condensed within
clouds forming at larger $z$ from the lifted disk gas which suffers
thermal instability and cools down to $T=10^4$ K (the minimum
temperature allowed in our simulations). These clouds are
characterized by large column densities, and are visible in Fig.
\ref{fig:colden} as red spots.

The cold, dense gas must eventually fall back onto the disk.  Figure
\ref{fig:vneg} shows the column density of the clouds with negative
vertical velocity (the cloud gas is selected as gas with $T\leq
5\times 10^4$ K).  This figure clearly demonstrates how the gas
lifted by the fountains falls back remaining mostly within the ring
of the disk in which the active area is located. This result
confirms what was found in Paper I and indicates that there is no
substantial radial dispersal of the metals from the Galactocentric
radius where they are produced. A similar result has been also
recently reported by \citet{spito08}.

Moreover, Figures \ref{fig:colden} and \ref{fig:vneg} show that the
small fraction of the rising gas escaping to the coronal sector
moves radially preferentially outward rather than inward. This is
due to  the decrease of the centripetal component of the gravity
with height and to the conservation of the angular momentum
\citep{shafi76,breg80}. The tendency of the MGF gas to move outward
is however, contrasted by its interaction with the hot halo (which
is assumed at rest). The fountain transfers part of its angular
momentum to the halo and thus tends to reduce its radial
displacement outward (see below).

\begin{figure}
\begin{center}
\psfig{figure=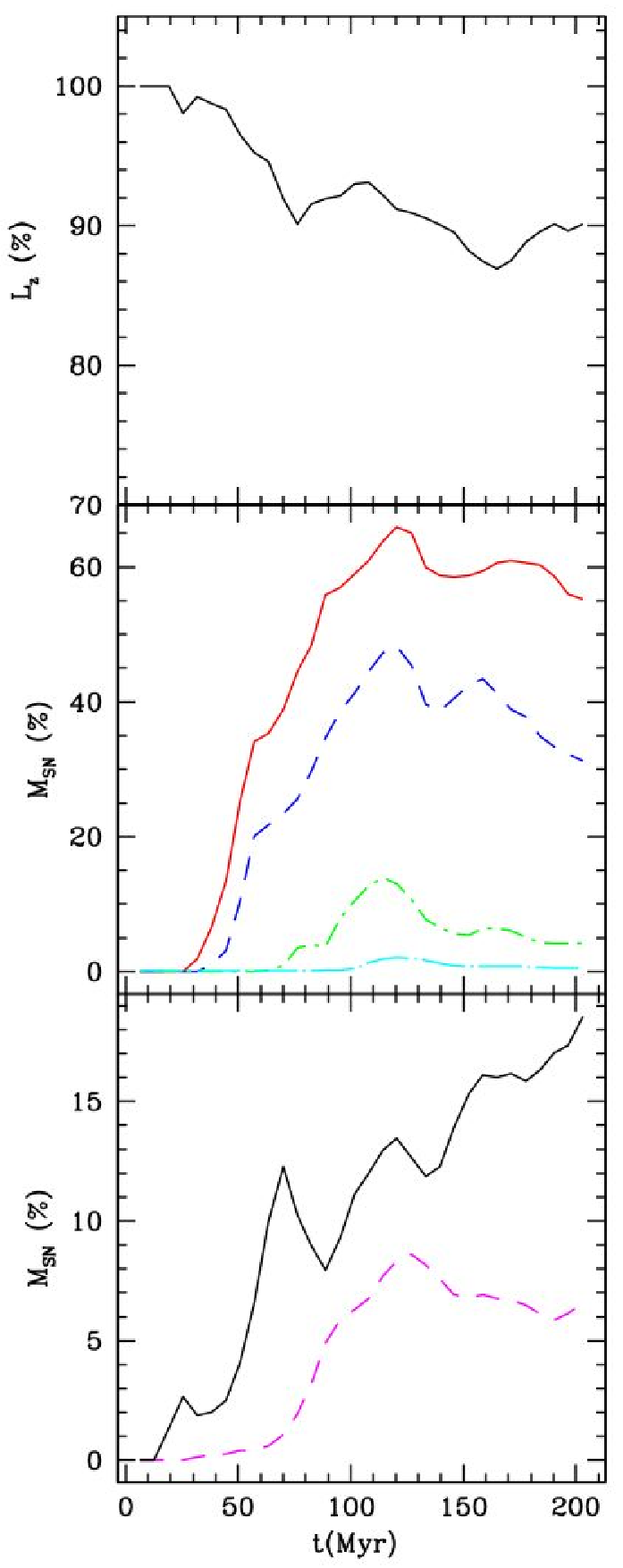,width=0.5\textwidth}
\end{center}
\caption{ Illustration of several quantities for the reference
  model. Top panel: angular momentum evolution of the SNII ejecta
  normalized to the angular momentum injected by the SN II
  ejecta. Middle panel: fraction of the mass of SN II ejecta located
  at $z>0.8$ kpc (solid line), $z>1.2$ kpc (dashed line), $z>2.4$ kpc
  (dot-dashed line) and $z>4$ kpc (long-dashed line). Bottom panel:
  ejecta mass fraction moving outside the active area toward the
  Galactic centre (dashed line) and toward the disk outskirt (solid
  line).  }
\label{fig:angmo}
\end{figure}

A more quantitative insight of the evolution of the gas circulation
is given by Fig.  \ref{fig:angmo}. The top panel shows the loss of
angular momentum of the SN II ejecta due to the interaction with the
halo gas \footnote{We point out that the time evolution of the
  relative angular momentum of $all$ the gas lifted by the fountains
  is quite similar to that relative to the SN II ejecta alone.}.  At
any time, the angular momentum displayed in the figure is normalized
to the total angular momentum injected by the SN II up to that time.
The middle panel illustrates the fraction of the SN II ejecta
located at different heights, while the bottom
panel shows the fraction of ejecta moving outside of the boundary of
the active area and testifies that the amount of gas moving outward
is larger than that moving inward. The latter starts to increase
after $\sim 70$ Myr, when a substantial amount of the lifted gas
starts to fall back (see section \ref{subsec:feedback}); having lost
part of its angular momentum, this gas tends to return at a
Galacocentric radius smaller than that at which it started its
journey. 

It is interesting to compare Fig. \ref{fig:angmo} with the analogous
figure in Paper I (Fig. 5) in order to highlight similarities and
differences with the SGF case. The relative loss of the angular
momentum of the fountain gas is essentially the same. This must be
expected as in MGFs a larger amount of fountain gas interacts with a
larger amount of halo gas, leaving the loss of specific angular
momentum essentially unaffected compared to SGFs. For analogous
reasons, also the relative amount of gas lifted above different
heights is similar; a larger amount of energy, in fact, moves a larger
amount of gas to similar altitudes. Obviously, as the SN explosions
keep going for a longer time, new gas persists at larger
heights, at variance with the SGF case.  Finally, we stress that in
the case of SGFs on the long term the fraction of gas moving outward
is smaller than that moving inward, at variance with the MGF
case. However, a glance to Fig. 5 of Paper I reveals that, initially,
also in the SGF case more gas moves outward rather than inward.  This
different behaviour is due to the fact that in the present model the
MGF flow is continuously supplied, while in the SGFs the SN engine
stops its activity after 30 Myr and the fountain gas, being no more
replenished with fresh angular momentum, is facilitated in its
centripetal displacement.

In any case, it is quite clear that most of the gas remains
essentially at the same Galactocentric distance where it has been
created.

\subsection{Cloud properties}
\label{subsec:cloud}

\begin{figure}
\begin{center}
\psfig{figure=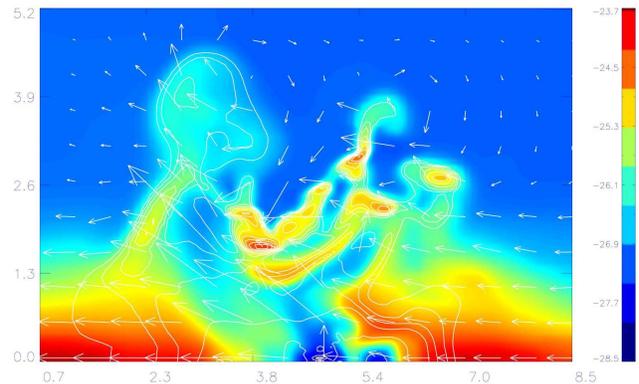,width=0.47\textwidth}
\end{center}
\caption{Distribution of the disk gas at $t=130$ Myr at the plane
  orthogonal to the Galactic disk and passing through the active area
  for the reference model. Isodensity curves of the ejecta are
  over-imposed, highlighting the fountain pattern and the cloud
  formation.  Since the density of the ejecta is much smaller than
  that of the ISM, it is represented in a different scale in order to
  be visible in the figure. The logarithmic density scale of the disk
  gas is given in g cm$^{-3}$ and the spatial scale along the $x$ and $z$ directions is in kpc.}
\label{fig:cloud}
\end{figure}

We now discuss in more detail the structure of the clouds forming
from the gas powered by the MGFs. In particular, the ISM lifted by
the fountains condenses essentially at the top of its trajectory,
and than falls back with velocities $\la$ 100 km s$^{-1}$.  $\sim
130$ Myr after the beginning of the simulation an equilibrium is
established between ascending and descending gas (cf. sect.
\ref{subsec:feedback}). After this time, the gas lifted above 1 kpc
has a mass $1.5\times 10^7$ $\msun$, 75\% of which
is condensed in dense filaments cooled down to $T=10^4$ K (see Fig.
\ref{fig:cloud}). The clouds are composed essentially of gas
originally from the disk ($\sim$96\%), and only a slight
contribution (4\%) from the halo gas that was compressed and cooled
after  interacting with the fountains.  As we have found in Paper I,
the SN ejecta plays a negligible role in the mass budget of the
fountains. After 130 Myr, $1.3\times 10^4$ SNe II have exploded and
delivered $2.1\times 10^5$ $\msun$ of ejecta, but only half of it
enters within the computational domain (the rest is injected in the
``lower'' half of the halo, which was not considered in our
simulations, for symmetry reasons, as remarked in Sec. 2). Nearly
50\% of the ejecta remains trapped within the disk, 35\% is
encompassed within the evolving clouds, and 15\% floats over the
disk as hot, diffuse gas. Thus the SN ejecta contribute for less
than 0.4\% to the clouds mass. For this reason, the chemical
composition of the clouds is unaffected by the metals produced by
the SNe II driving the fountains. As each supernova delivers on
average 3 $\msun$ of metals \citep[cf.][]{mar07}, a total mass of
$2\times 10^4$ $\msun$ of heavy elements is ejected by the fountain
within the computational domain.  As a result, the metallicity
increment in the clouds due to the freshly delivered metals is
$\Delta Z=5.7\times 10^{-4}$, rather negligible compared to the
solar metallicity of the ISM. This conclusion holds also if one
considers the contribution of SNe II belonging to OB associations
hosting less than 30 SN progenitors, and representing half of the
total number of supernovae (see section \ref{subsubsec:mltas}).

In conclusion, the clouds forming in our simulation have solar
metallicities and velocities lower than 100 km s$^{-1}$, and are
therefore likely to be associated to IVCs rather than HVCs
\citep{wak08}. This result is in agreement with that found in Paper I
when considering single fountains.

\subsection{Disk-halo energy exchange}
\label{subsec:feedback}

\begin{figure}
\begin{center}
\psfig{figure=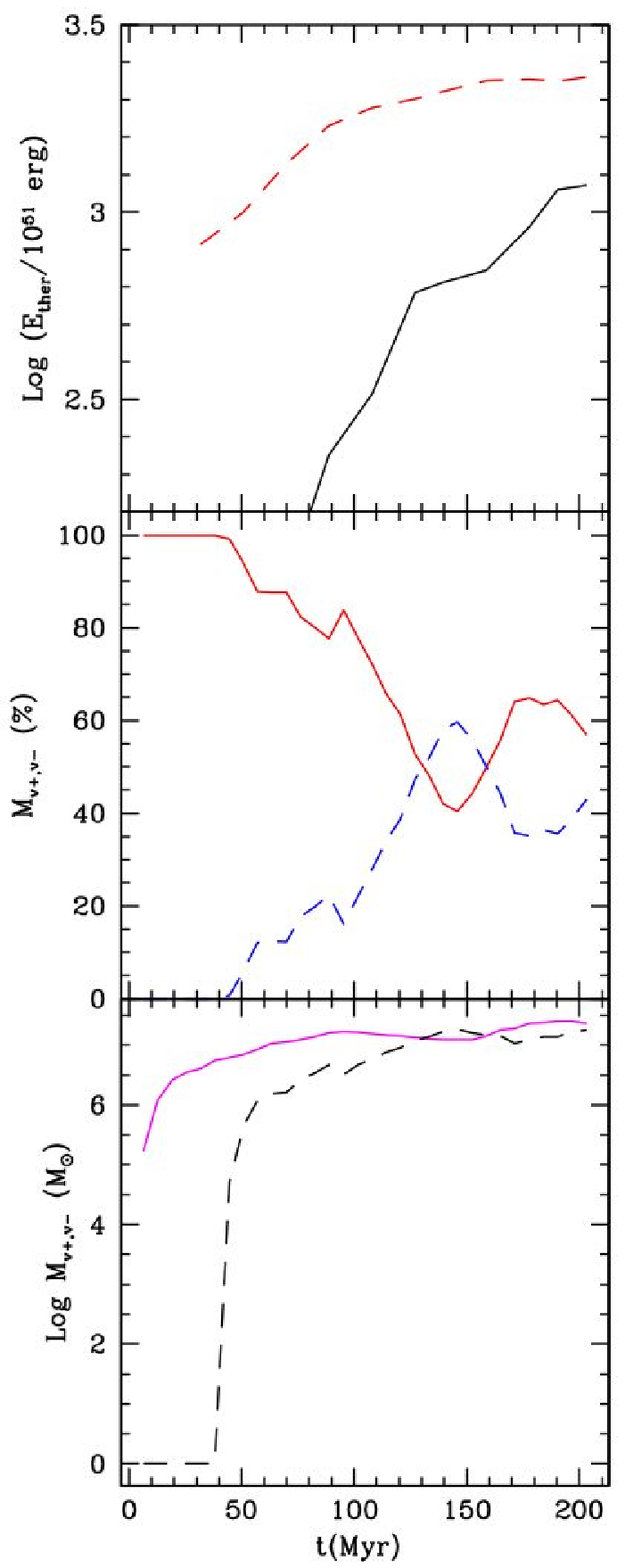,width=0.5\textwidth}
\end{center}
\caption{Illustration of several quantities of the reference
  model. Top panel: amount of thermal energy of the disk gas
  transferred to the gaseous halo (solid line), thermal energy lost by
  the halo gas condensed onto the clouds (dashed line); see text for
  details. Middle panel: fraction of the total amount of gas that is
  rising (solid line) and that is descending (dashed line). Bottom
  panel: the same as in the middle panel, but with the absolute values
  of the amounts of gas.  }
\label{fig:equil}
\end{figure}

Other details of the gas flows of the reference model are given in
Fig. \ref{fig:equil}, and are intended to shed light on the
feedback between the disk and the hot halo. This
energy exchange regulates the evolution of the halo gas and influences
the star formation history in the disk. It is also relevant in the related
issue of the so-called ``over-cooling'' problem in cosmological
simulations of galaxy formation, which generally overestimate the amount of radiatively
cooled gas in galaxies. Energy feedback from SNe and/or AGN, albeit poorly
understood, is thought to be a crucial ingredient to reconcile hierarchical
models of galaxy formation with the observations \citep[e.g.][]{nav97,crot06}.
Supernovae alone are probably incapable to fully solve the problem \citep{tor04},
but is nevertheless important to quantify the SN energy which heats the hot halo gas
to correctly estimate its cooling rate and the disk-halo mass exchange.

As in our model the amount of thermal energy of all the halo is much
larger than the energy injected by the SNe, a direct evaluation of its
variation due to the stellar explosions would be rather inaccurate
(also because of the perturbations generated at the grid boundaries,
as discussed in Paper I). We thereby check the energy balance of the
halo-fountains interaction as follows. In the top panel of
Fig. \ref{fig:equil} we report the thermal energy of the disk gas that
is lifted by the fountains and not condensed into clouds. This
rarefied gas endures over the disk and its thermal energy is
incorporated in the hot halo. On the other hand, some coronal gas
cools and condenses over the clouds created by the MGFs.  The thermal
energy radiated by this cold gas is also shown in the top panel of
Fig. \ref{fig:equil}. At the end of the simulation, the halo has gained
an energy twice larger than that radiated due to the
interaction with the fountains, with a net increase of thermal energy
of $\sim 10$\% of the energy released by all the SNe.  This value
for the SNe heating efficiency is
expected to remain stable once an approximately steady regime is
established between rising and falling gas.

The achievement of this regime is visible in the middle panel of
Fig. \ref{fig:equil} that shows the fraction of rising (solid line)
and descending (dashed line) fountain gas.  At early times only the
rising gas is present. After nearly 40 Myr the first
clouds form and fall back. A dynamical equilibrium between ascending
and descending gas is established after 130 Myr. The same result is
illustrated in the bottom panel in terms of the absolute values of
the rising and falling masses.

As pointed out, we cannot make an exact evaluation of the energy
balance between the fountains and the coronal gas.  In the analysis
above (see the top panel of Fig. 7), we neglected the gain
of the kinetic energy of the hot gas interacting with the galactic fountains
and the change in the potential energy of the whole halo.
Also, we neglected
the radiative losses of the hot gas, possibly enhanced by
the compression due to the fountain expansion. 
However, the flow of the halo gas remains
subsonic and its cooling time is longer than the time of the
simulation. We thus  conclude that a net energy gain is attained by
the halo during its interaction with the MGFs.

\section{Model at R=4.5 kpc}
 \label{sec:mr45mod}

 In this section we explore the influence of the Galactocentric
 distance on the gasdynamics of the MGFs. In particular, we have run a
 model similar to the RM, but with the active area centred
 at $R=4.5$ kpc. Here the gas of the disk is denser and has a larger
 effective vertical scale heigh $H_{\rm eff}$. As a consequence, the disk-halo
 transition occurs at $z=1.35$ kpc, instead of $z=0.8$ kpc as in the
 RM (see also Paper I).

\begin{figure*}
\begin{center}
\psfig{figure=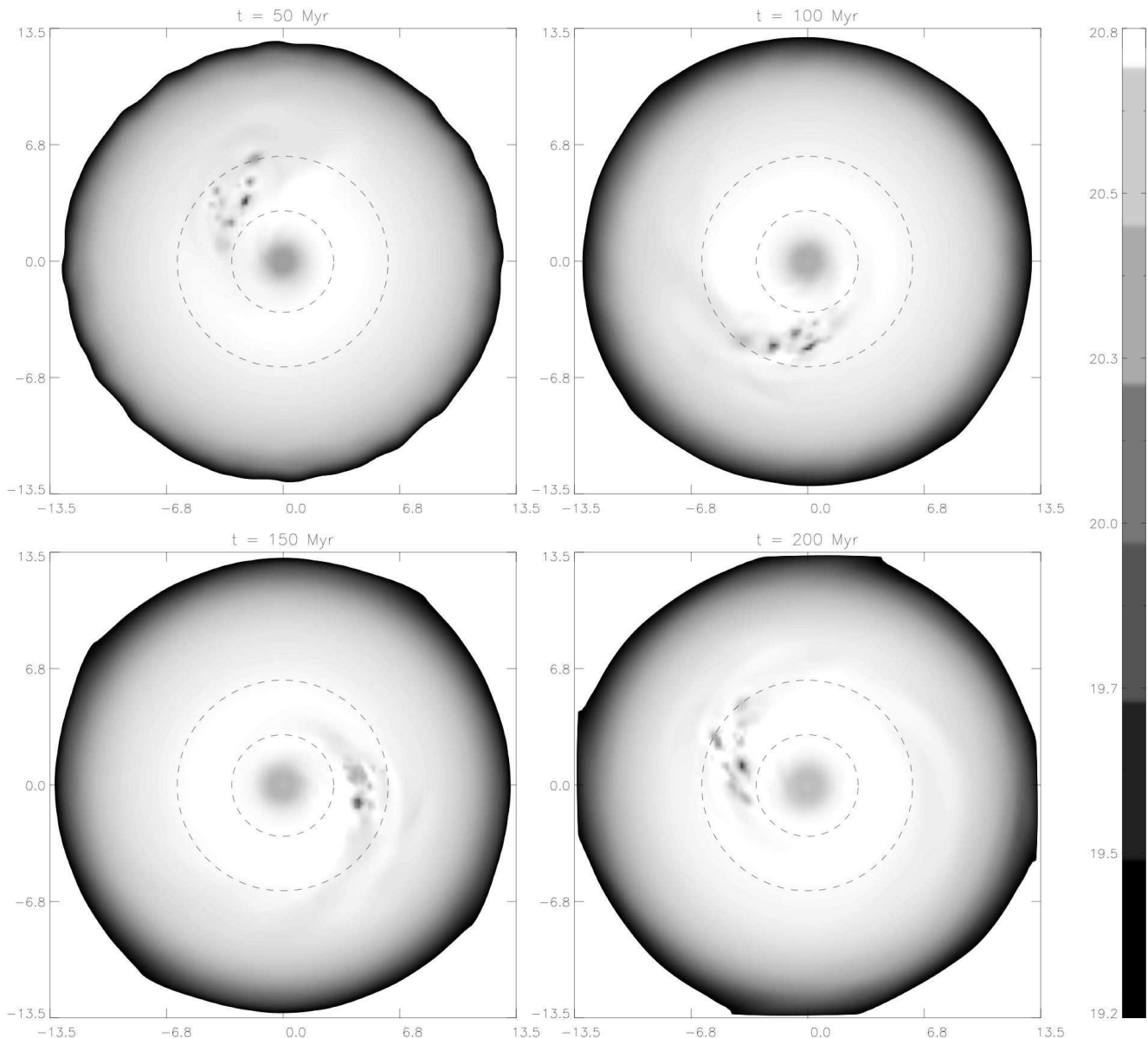,width=1.0\textwidth}
\end{center}
\caption{ The same as in  Fig. \ref{fig:coldenism}, but for the
Galactocentric distance $R=4.5$ kpc. } \label{fig:coldenism4}
\end{figure*}

The larger density of the ambient medium makes more difficult the
expansion of the SN bubbles through the plane and accelerates their
backfilling. The reduced size and lifetime of the holes also reduce
their probability of merging.  This is apparent in Fig.
\ref{fig:coldenism4}, where the number and the size of the holes are
smaller than in the reference model (cf. Fig. \ref{fig:coldenism}).

\begin{figure*}
\begin{center}
\psfig{figure=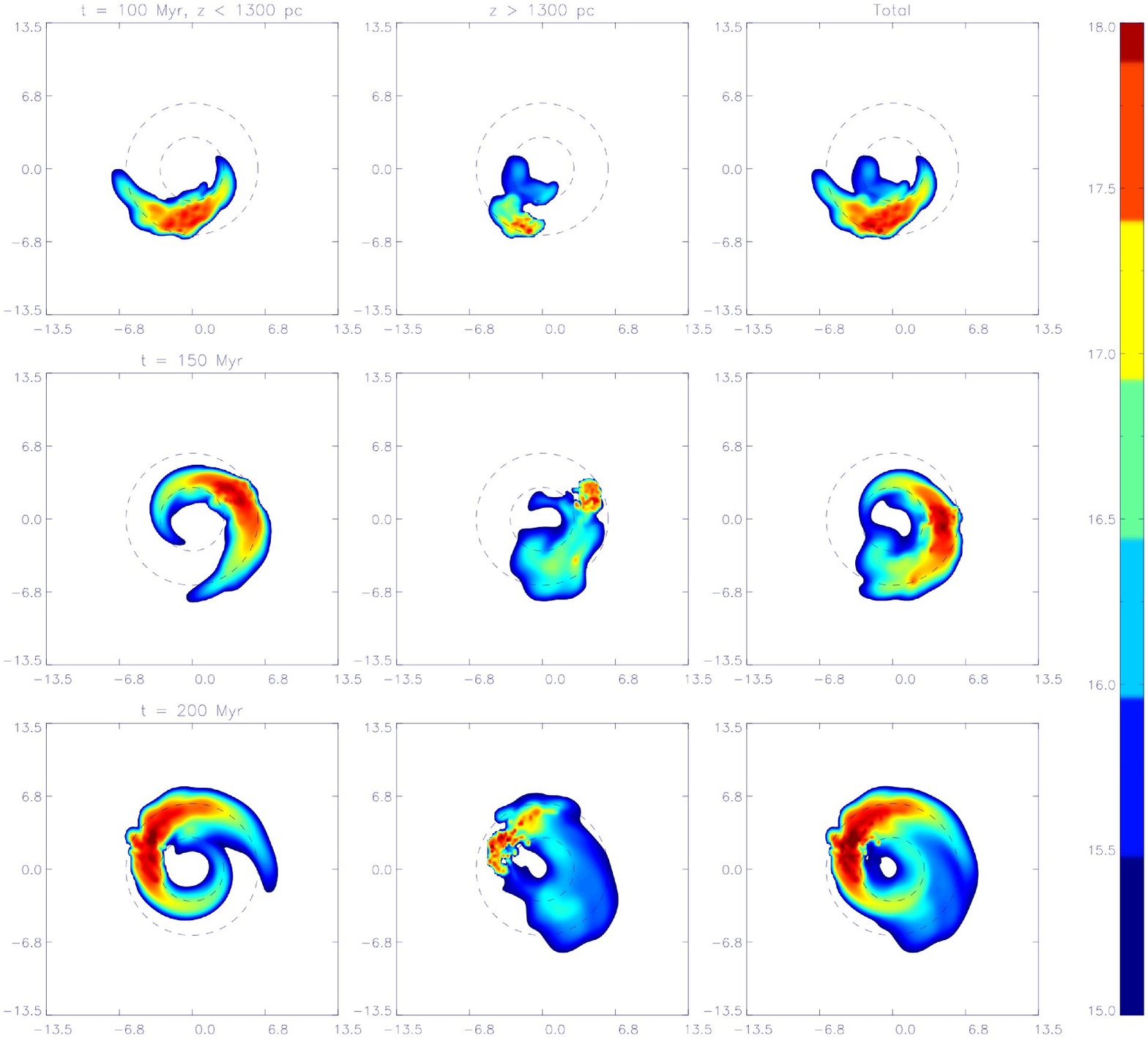,width=1.0\textwidth}
\end{center}
\caption{ The same as in Fig. \ref{fig:colden}, but for the
  Galactocentric distance $R=4.5$ kpc. In this case the left and
  central panels give the column densities calculated in the intervals
  $z<1.3$ kpc and $z>1.3$ kpc, respectively, since at this
  Galactocentric distance the disk/halo transition occurs at $z=1.3$
  kpc.  }
\label{fig:colden4}
\end{figure*}

In Paper I it was shown how the vertical distribution of the disk
gas influences the dynamics of a SGF and the shape of the
chimney carved by the fountain.  In particular, a single fountain
located at $R=4.5$ kpc gives rise to a sort of well collimated
outflow that  crosses the disk/halo transition only barely. On the
contrary, given the lower density and scale height of the local gas,
the same fountain located at $R=8.5$ kpc experiences a well defined
break out. In the case of MGFs these differences are greatly reduced
because the energy powering the gas flow in this case is much larger
than the critical luminosity $L_{\rm cr}$ at any Galactocentric
distance.

\begin{figure*}
\begin{center}
\psfig{figure=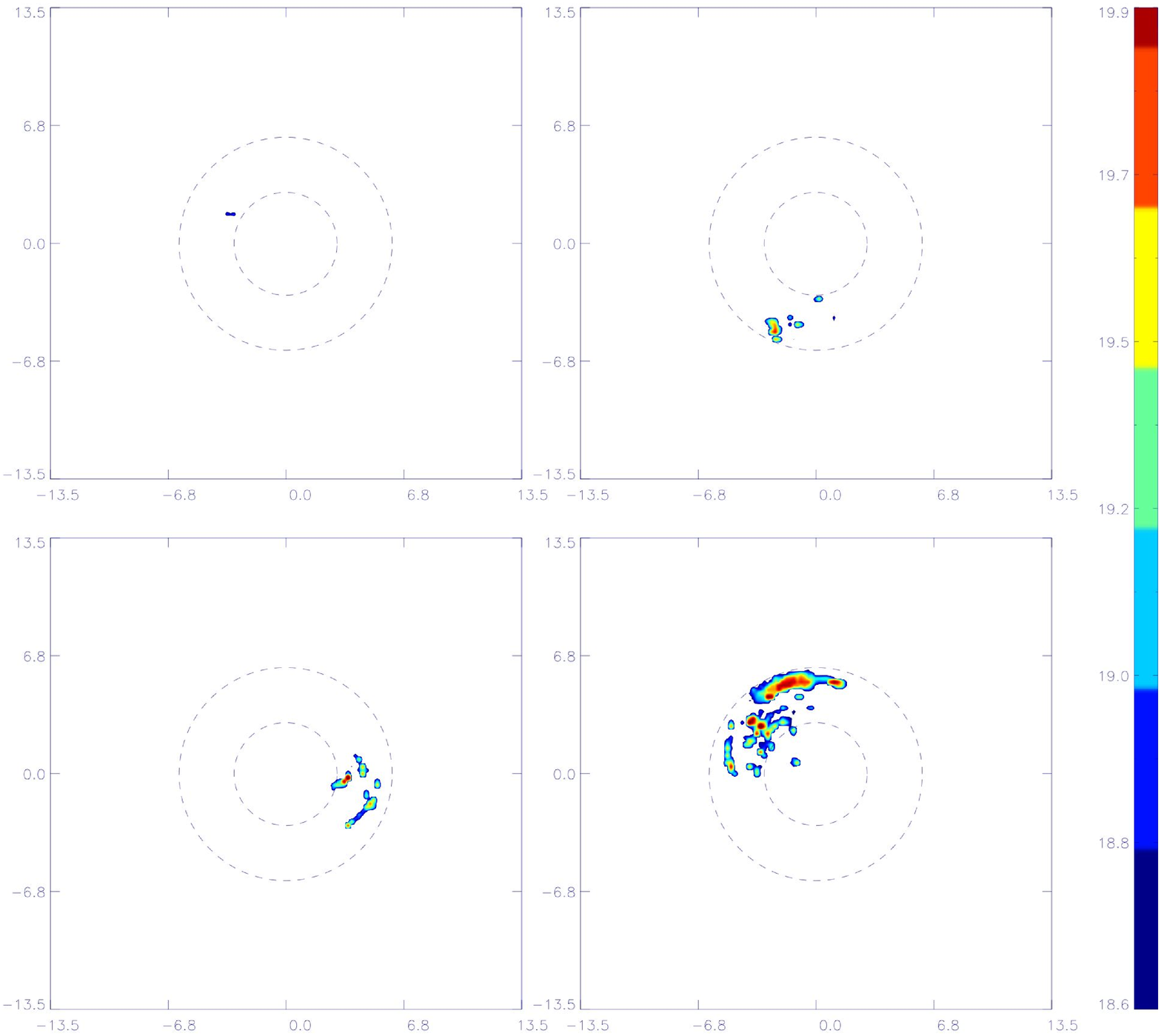,width=1.0\textwidth}
\end{center}
\caption{ The same as in Fig. \ref{fig:vneg}, but for the
galactocentric distance $R=4.5$ kpc } \label{fig:vneg4}
\end{figure*}

A somewhat larger alignment is still present at $R=4.5$ kpc also in
the case of MGFs, as apparent when one compares Fig.
\ref{fig:colden} with  Fig. \ref{fig:colden4} (where the different
position of the disk/halo transition has been taken into account).
The amount of diffuse gas spread outside the circular sector is
smaller than in the RM. The same conclusion can be reached looking
at Fig. \ref{fig:vneg4}. The falling gas is clearly less radially
dispersed compared to the reference model (cf. Fig. \ref{fig:vneg}).
Note that, since in the present model the circumference covered by
the active area is shorter, after 200 Myr a complete ring is
described in all the panels of the bottom raw of Fig.
\ref{fig:vneg4}.

\begin{figure}
\begin{center}
\psfig{figure=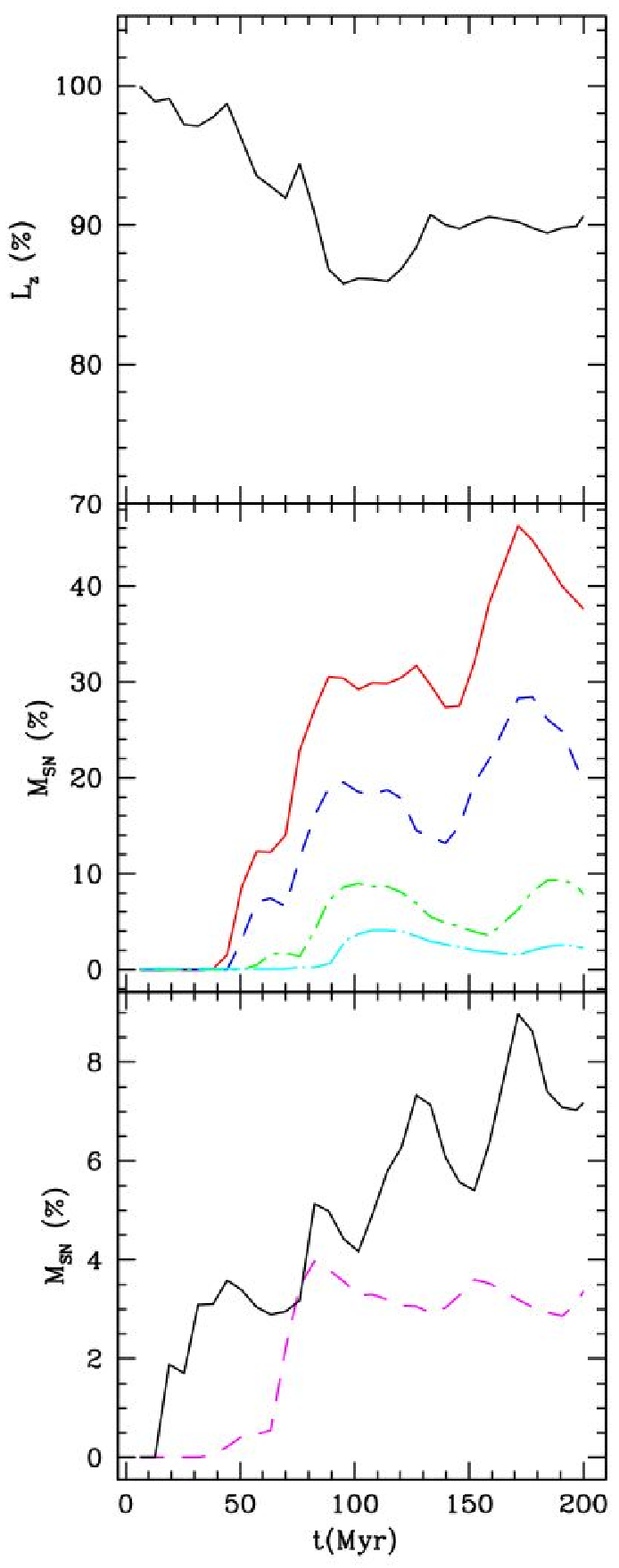,width=0.5\textwidth}
\end{center}
\caption{ The same as in Fig. \ref{fig:angmo}, but for the model with
  the active area centred at the Galactocentric radius $R=4.5$ kpc.}
\label{fig:angmo4}
\end{figure}

As for the RM, more quantitative information can be obtained looking
at the Fig. \ref{fig:angmo4}. The loss of angular momentum is
essentially the same as in the RM (top panel). Comparing to the SGF
evolution (Fig. 9 in Paper I), we remark that in this latter case the
angular momentum, after an initial drop, slowly goes up again. As
discussed in Paper I, this is due to the fact that a larger fraction
of the SGF gas falls back onto the thicker disk gaining new angular
momentum. This effect is not present in the MGF model because,
as stressed above, given the larger energy of the fountain, the disk
thickness does not play a significant role

The vertical extent covered by the gas (middle panel of
Fig. \ref{fig:angmo4}) is somewhat lower than in the RM as the density
and the scale height of the disk are larger in this case. The
differences with the SGF case are analogous to those discussed for the
RM, and do not repeat them here.

The smaller radial dispersion of the MGF gas emphasized above is
illustrated in the bottom panel of Fig. \ref{fig:angmo4} showing that
the amount of gas moving radially outside the active area in nearly
halved compared to the RM.

\begin{figure}
\begin{center}
\psfig{figure=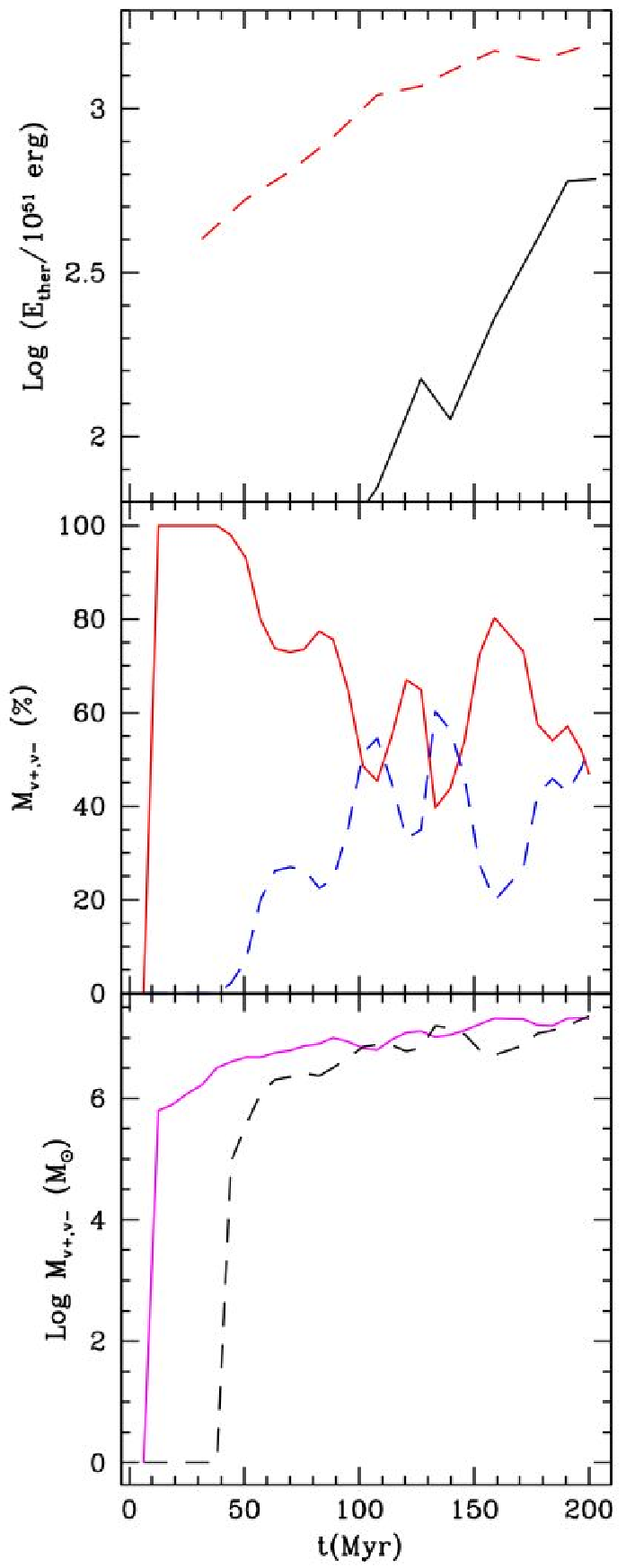,width=0.49\textwidth}
\end{center}
\caption{ the same as in Fig. \ref{fig:equil}, but for the model with
  the active area centred at the Galactocentric radius $R=4$ kpc. }
\label{fig:equil4}
\end{figure}

Finally, from the top panel of Fig. \ref{fig:equil4} we see that in
this model the halo gas acquires  10\% of the SN energy at the end
of the simulation, i.e., the same amount as in the RM. Also the time
at which the quantity of ascending and descending gas become
comparable is similar to that of the RM (middle and bottom panels of
Fig. \ref{fig:equil4}).

\section{Interaction with accreting gas}
\label{sec:accr}

As pointed out in Sec. 1, the gas flow of the fountains is likely to
sweep up cold gas, but it is unclear whether this gas arises from cooling
of the hot halo or from cold streams falling from large distances.
\citet{frabi08} argue that this latter possibility is the most
realistic. If the cold falling gas is clumpy, its clouds cannot be
more massive than $10^7$ $\msun$, otherwise the solar neighbourhood
would be heated by their passage \citep{laos85,totos92}. Actually,
while HVCs surrounding our Galaxy have masses of order $10^{6-7}$
$\msun$ \citep[e.g.][]{wak08}, deep observations around M31 reveal
the presence of numerous HVCs with masses down to $10^4$ $\msun$
\citep{west05}.

The study of the possible accretion of gas onto galaxies is
important for several reasons.  It is generally believed that
accretion must replace the gas consumed by star formation which has
been remarkably constant in the solar neighbourhood over the Galaxy
lifetime \citep[e.g.][]{bidebe00}.  Another reason for studying the
galactic accretion is linked to the mounting evidence of massive
haloes of neutral and ionized gas surrounding nearby spiral galaxies
\citep[e.g.][]{safra08}.  These thick disks rotate more slowly than
the thin disks and show inflow motions. \citet{frabi08} have shown
that these haloes cannot be sustained only by galactic fountains.
According to them, the  gas from the fountains interacting with a
pre-existing hot corona would make it to co-rotate with the disk
after a very short time ($\la 1$ Myr) which is not consistent with
the observations. This result led them to conclude  that a
substantial accretion of low angular momentum material from the
intergalactic medium would be required in order to assure a slower
rotating halo.

In the model of \citet{frabi08} the clouds are treated
as bullets moving ballistically, and hydrodynamical effects are absent
or only roughly represented. In order to investigate the issues above
more realistically from an hydrodynamical point of view, we have
simulated the interaction of the MGFs with infalling gas considering
two different models, one in which the gas infall is described as a
continuous drizzle, and another in which a single, denser cloud is
assumed to fall toward the active area.

\subsection{The ``drizzle'' model}
\label{subsection:drizzle}

This model is similar to the reference model, but we assume that a
gas of density $n=10^{-3}$ cm$^{-3}$ drops from the ``top'' boundary
of the computational grid with a vertical velocity $v_{\rm z}=-150$
km s$^{-1}$. Such a flux extended over all the Galactic disk would
correspond to an accretion rate of 5.2 $\msun$ yr$^{-1}$ (from both
sides). This is much higher than the accretion rate $\dot M\sim 0.2$
$\msun$ yr$^{-1}$ associated with the HVCs, and also higher than the
rate of $\sim 2$ $\msun$ yr$^{-1}$ needed to sustain the star
formation in the solar neighbourhood.  However, the accretion rate we
assume is comparable to that in \citet{frabi08} (which was adopted
for the case of NGC 891) and this will allow a direct comparison
with their work.

As we consider the fountain activity only in a limited region of the
disk plane, we let the rain fall from the top boundary only over the
ring sector on the disk where the active area moves (cf. Fig.
\ref{fig:coldenism}). We have considered only this  region because
 most of the gas set in motion by the fountains remains within this
sector (as we have seen in section \ref{sec:mrefmod}), so that  all
the features of the interaction of the fountains with the rain can
be captured.

\begin{figure}
\begin{center}
\psfig{figure=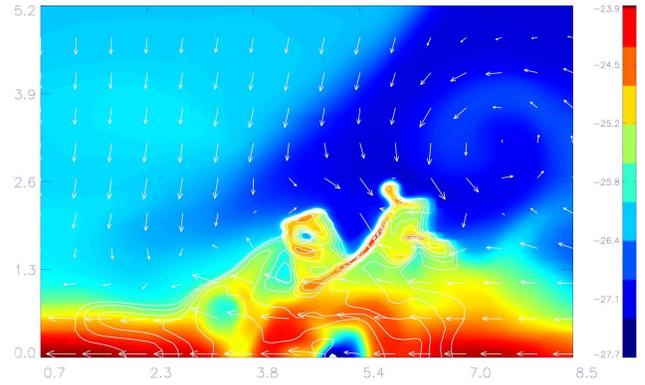,width=0.47\textwidth}
\end{center}
\caption{The same as in  Fig. \ref{fig:cloud}, but for a model
considering a continuous external gas infall over the active area
(the drizzle model).} \label{fig:rain}
\end{figure}

As the infalling gas interacts with the hot halo and with the
ascending gas of the fountains, vortices of $\sim 2.5$ kpc form and
the density of the gas above the disk fluctuates, as shown in Fig.
\ref{fig:rain}. These vortices are likely an artifact of the method
we use to inject the accreting gas. However, these features do not alter
the flow generated by the fountains, being the vortices energetically
negligible compared to the energy injected by the SNe.
Figure \ref{fig:rain} corresponds to the same time and grid
slice of the snapshot depicted in the RM of Fig. \ref{fig:cloud}.
Comparing both models,  we see that in the drizzle model the disk
gas set in motion by the fountains reaches lower heights and tends
to lag behind the active area (toward the right hand side in the
figure).

\begin{figure}
\begin{center}
\psfig{figure=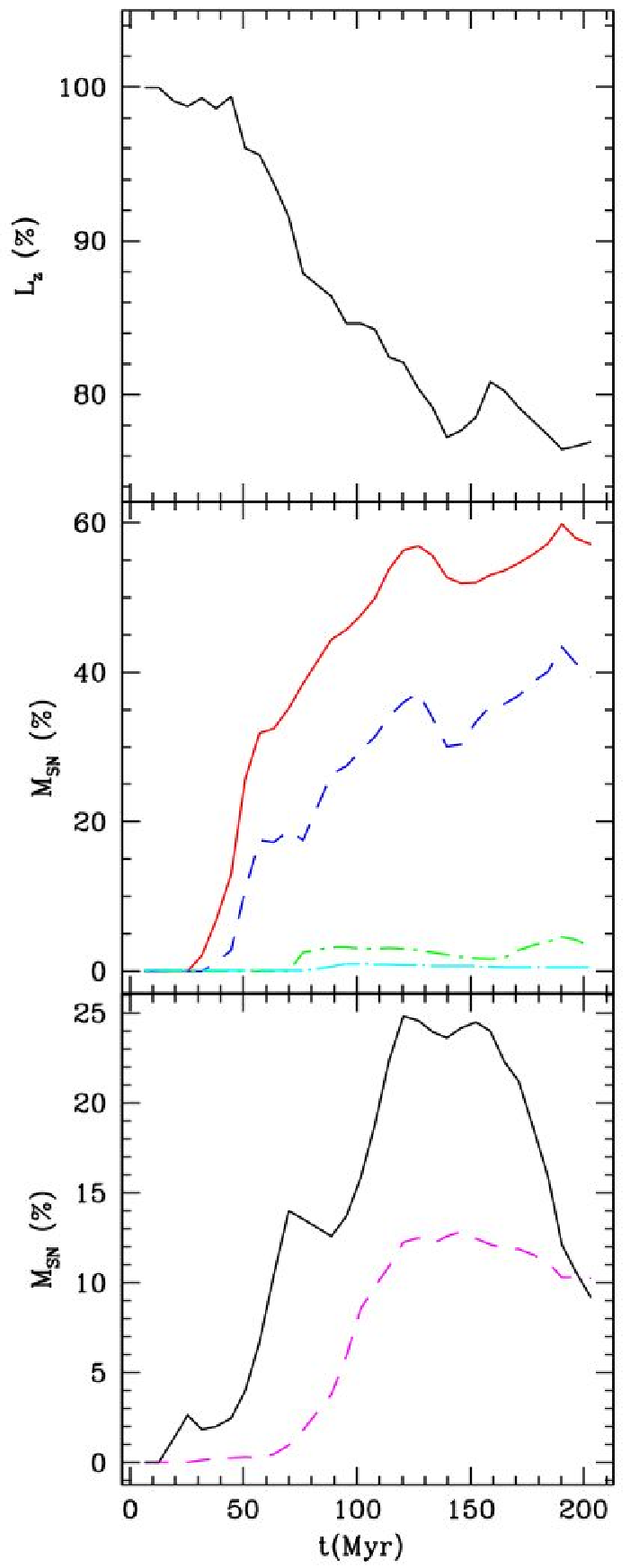,width=0.5\textwidth}
\end{center}
\caption{ The same as in  Fig. \ref{fig:angmo}, but for the drizzle
model.} \label{fig:angmor}
\end{figure}

A more quantitative analysis is given in Fig. \ref{fig:angmor}. The
angular momentum loss (top panel) is more than doubled relative to
the RM. This explains the larger lag of the fountain flow relative
to the disk, as noted above. The middle panel quantifies the
vertical distribution of the fountain gas; it is essentially similar
to the reference model up to $z=1.2$ kpc (cf. Fig. \ref{fig:angmo}),
but less gas is found above $z=2.4$ kpc.

Once an overall balance is reached between ascending and descending
gas, as in the RM, 25\% of the lifted gas remains rather diffuse,
while 75\% of it condenses in dense filaments. The total mass of
these clouds is $7\times 10^6$ $\msun$, i.e.,  nearly 60\% of what
was found in the RM; such reduction is related to the lower vertical
excursion of the clouds and their shorter ``lifetime''. It is
interesting to analyze the ``composition'' of the clouds. As in the
RM, the contribution to their mass by the SN ejecta amounts to a
negligible 0.3\%. This time, however, the amount of disk gas is
lower (87.9\%), while the fraction of ``external'' gas is three
times larger (11.8\%). This result is obviously due to the addition
of the rain material to the halo gas. This denser ambient medium
cools more easily when it interacts with the clouds and condenses
onto them more efficiently.

The accretion mass onto the clouds is a crucial ingredient in the
model of \citet{frabi08}. In their model, they integrated the orbits
of the clouds that are ejected from the disk of a spiral galaxy and
 move up ballistically to the halo and then return to the disk. As
the clouds accrete mass from  external infalling gas, their velocity
decreases due to momentum conservation. These authors have then
shown that, if  the thick disk of neutral gas of the spiral galaxies
is made up of clouds lifted by galactic fountains,  the accretion of
external gas with low angular momentum could explain why the
observed rotational curve of these disk galaxies decreases with $z$.
It is interesting to compare our results with those found by
\citet{frabi08}.

In order to evaluate the mass accretion onto our clouds, let us define
$m_0$ as the ``initial mass of the clouds'' (i.e., the mass of the
disk gas lifted by the fountains that has condensed into clouds), and
$\delta m$ as the amount of ambient gas (i.e. the gas of the halo and
the drizzle) that is swept by the clouds and increase their mass up to
$m=m_0\delta m$. From our simulations, we obtain $\delta m/m_0=0.13$;
this result is in reasonable agreement with what found by
\citet{frabi08} in the case of NGC 2403 (cf. their Fig. 6).

\begin{figure}
\begin{center}
\psfig{figure=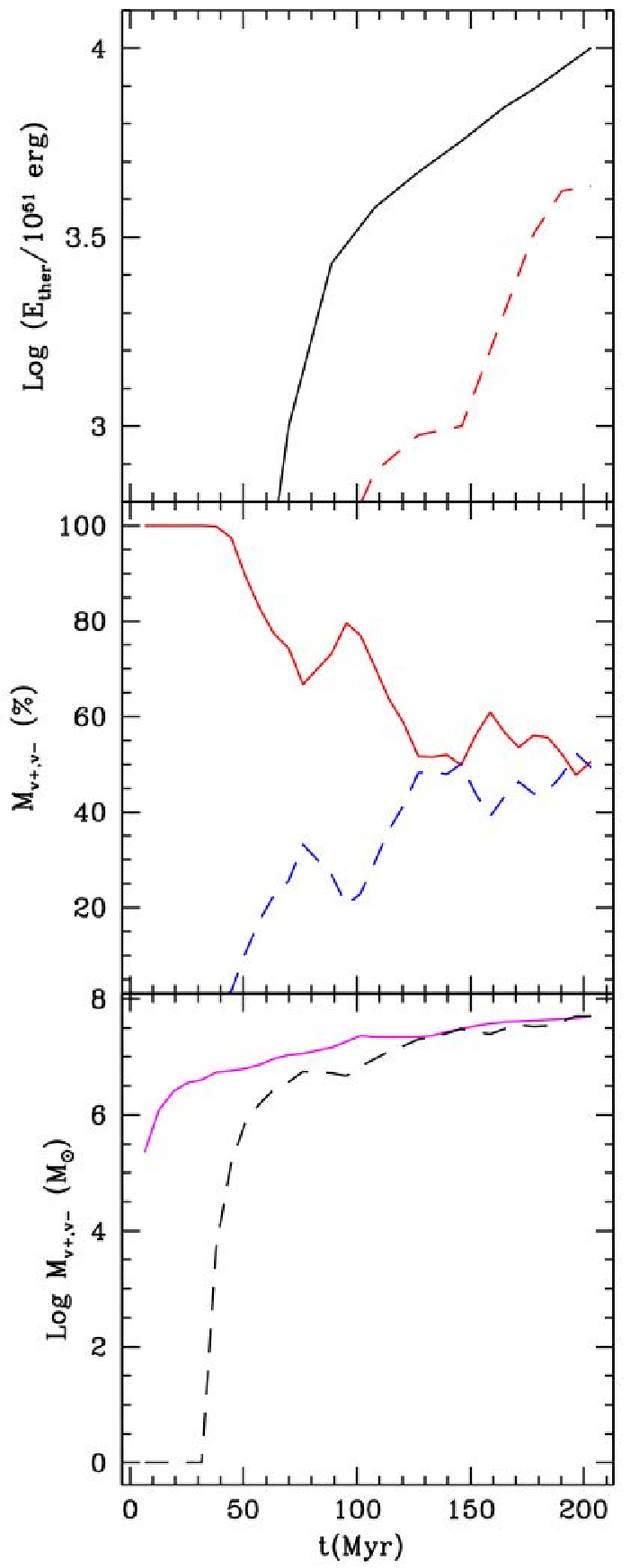,width=0.5\textwidth}
\end{center}
\caption{ The same as in  Fig. \ref{fig:equil}, but for the drizzle
model.} \label{fig:equilr}
\end{figure}

We can pursuit further the analysis above computing the specific
accretion rate $\alpha=\dot m/m$. To this end we must evaluate the
clouds' lifetime $t_{\rm lf}$ in order to work out $\alpha=\delta
m/(m_0t_{\rm lf})$. From Fig. \ref{fig:equilr} we note that ascending
and descending gas reach a dynamical equilibrium after $t\ga 70$
Myr. This is thus the characteristic time spent by a cloud during its
journey, we obtain $\alpha\sim 1.9$ Gyr$^{-1}$, which is in good
agreement with the value $\alpha\sim 2$ Gyr$^{-1}$ found by
\citet{frabi08}. Although this agreement must not be super-estimated
given the approximations involved in the previous calculations, yet we
consider it significant.

\begin{figure}
\begin{center}
\psfig{figure=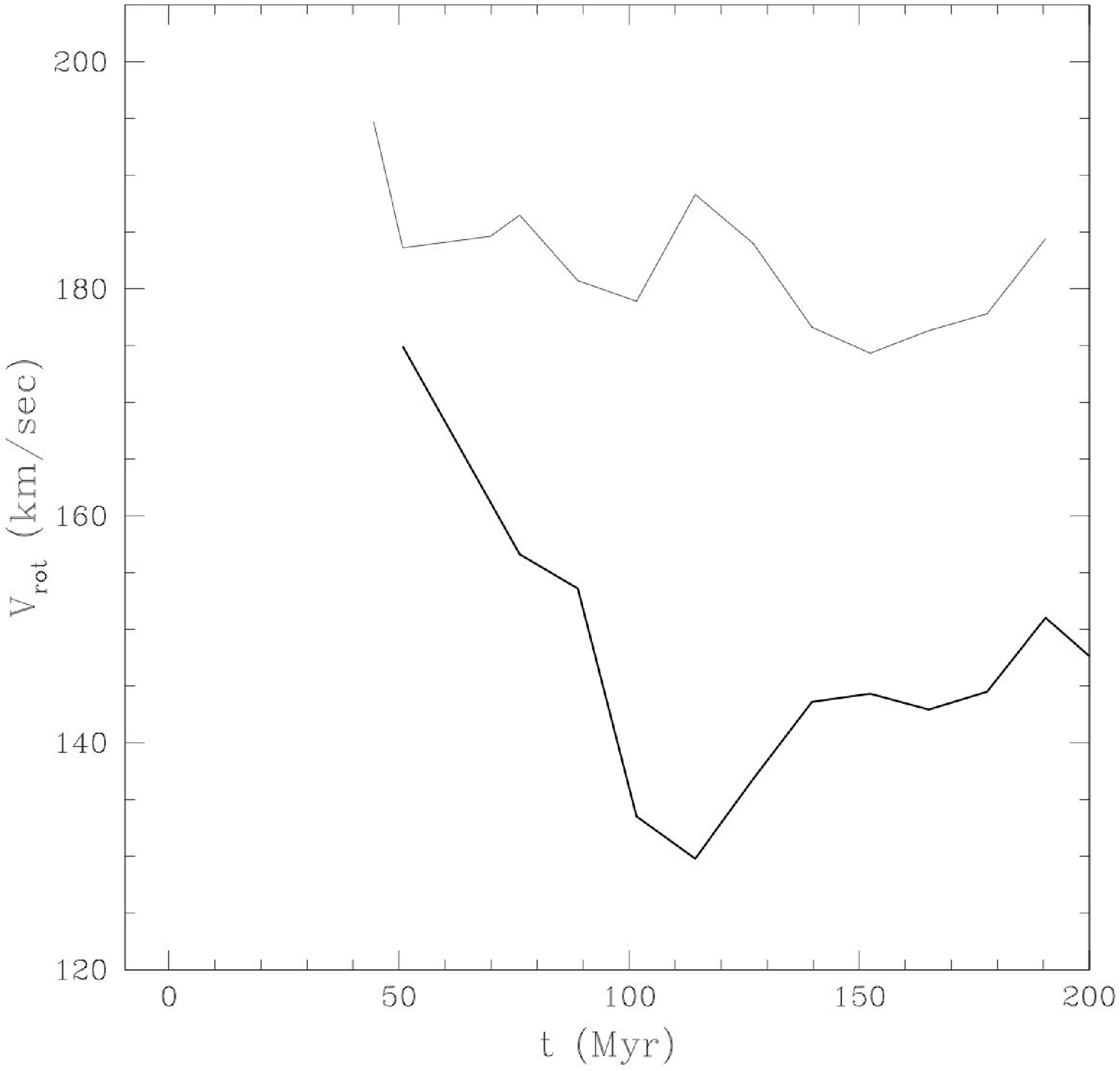,width=0.5\textwidth}
\end{center}
\caption{Time evolution of of the circular velocities of the clouds at
   $z=1.8$ kpc for the reference model (thin line) and the drizzle
  model (thick line).  }
\label{fig:vhalo}
\end{figure}

We now evaluate how the external mass accretion onto the clouds
affects their rotational velocity.  Obviously, given the limited
extension of the active area in our model, we cannot build a thick
disk of neutral hydrogen out of it, nor reproduce the entire disk
rotational curve. However, some interesting conclusions can be drawn
from the simulations above. The time evolution of the rotational
velocities at $z=1.8$ kpc are illustrated in the top panel of Fig.
\ref{fig:vhalo} for both the reference and the drizzle models. They
give the mass weighed mean velocity of a ''slice'' of clouds at
$z=1.8$ kpc. Likewise the angular momentum (cf. the top panels of
Figs. \ref{fig:angmo} and \ref{fig:angmor}), the values of the
velocities tend to stabilize after an initial drop.

It is an easy matter to verify whether the difference in the cloud
rotation velocity between the drizzle model and the RM is entirely
due to the difference in the amount of mass increment (as suggested
by the ballistic model of Fraternalli \& Binney 2008),  or not. In
that case we would have
\begin{equation}
{v'_{\rm rot}\over v_{\rm rot}}={{1+{\delta m \over m_0}}\over
{1+{\delta m' \over m'_0}}},
\end{equation}
\noindent where the prime quantities refer to the drizzle model.
From Fig. \ref{fig:vhalo} we find that at $t=200$ Myr, $v'_{\rm
rot}/ v_{\rm rot}=0.83$. On the other hand,  computing  the amount
of accreted matter onto the clouds from the simulations we find that
the right hand side of equation (4) gives a larger ratio $v'_{\rm
rot}/ v_{\rm rot}=0.92$.  This means that only  part of the rotation
velocity drop is due to the accretion of external material to the
clouds.  A rough estimate of the total momentum flux of the system
(that includes both the clouds momentum flux and the halo gas and
ram pressures) indicates that the remaining drop of the rotation
velocity of the clouds is due to their interaction with the hot halo
which in the case of the drizzle model is much denser due to the
external accretion.

In conclusion, though our simulations are restricted to a small area
of the Galactic disk, and therefore are not suitable to reproduce
the rotational curve of the entire thick disk or of the hot halo,
they seem to indicate that the presence of an external gas infall
may help to slow down the rotation of the gas in the clouds (see
Fig. 16) and thus the amount of angular momentum that is transferred
to the coronal gas, as suggested by \citet{frabi08}.

A final remark is in order with regard to the discussion above.  In
the model of \citet{frabi08} the clouds are described as pellets
that preserve their identity during all their journey until the
return to the disk. Let us define $\chi$ as the ratio between the
density of the cloud and that of the hot halo, and the flow time as
$t_{\rm flow}=D/v$, where $D$ and $v$ are the cloud diameter and
velocity, respectively. \citet{frabi08} estimate that typically
$\chi\sim 300$, $D=100$ pc and $v=100$ km s$^{-1}$. With these
figures, it turns out that the orbital time of a cloud is of the
same order of the drag time $t_{\rm
  drag}=\chi t_{\rm flow}\sim 300$ Myr. However, a
cloud can be destroyed by Kelvin-Helmholtz (K-H) instabilities in a
time scale $t_{\rm KH}=\chi ^{0.5}t_{\rm flow}\sim 17$ Myr. Thus, in
the model of \citet{frabi08} the cloud material is likely to linger
above the disk as diluted gas. In our models, instead, the clouds
never reach very large densities, and for the densest structures we
obtain $\chi \sim 100$ and $D\sim 500$ pc. As a consequence, it is
$t_{\rm KH}\sim 50$ Myr, which is of the same order of the cloud
lifetime. Simulations with higher spatial resolution would be
required in order to follow appropriately how the cloud evolution is
influenced by K-H instabilities. Such more demanding  simulations are
planned in a near future.


\subsection{The ``cloud'' model}
\label{subsection:cloudm}

\begin{figure*}
\begin{center}
\psfig{figure=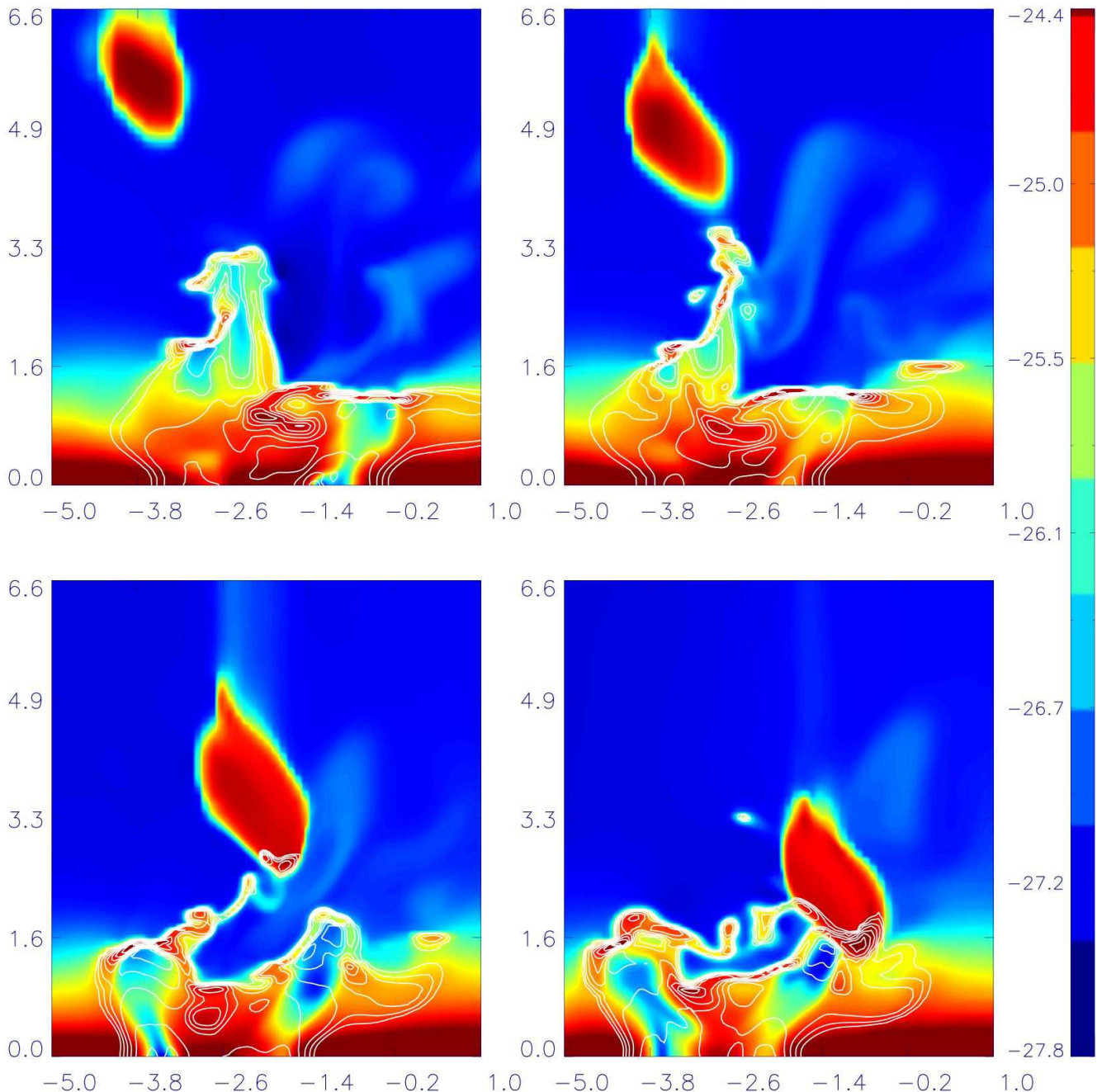,width=1\textwidth}
\end{center}
\caption{The same as in Fig. \ref{fig:cloud}, but for the cloud model,
  where an external cloud falls over the MGFs.  The four panels
  represent snapshots taken at $t=130$ Myr (top-left), $t=136.5$ Myr
  (top-right),$t=143$ Myr (bottom-left),$t=149.5$ Myr (bottom-right).
  As the Galaxy rotates, the falling cloud appears to move rightward.}
\label{fig:rainc}
\end{figure*}

In this section we study the effect of an external cloud falling toward
the galaxy on the fountain gas dynamics. We do not follow the subsequent interaction
between the cloud and the disk gas \citep[see][and references therein for a review on this subject]{jb04}.
We consider a cloud entering the grid with a velocity $v_{\rm z}=-150$
km s$^{-1}$ from the top boundary, with mass
$M=3.5\times 10^6$ $\msun$, which is consistent with the estimate
for Complex C \citep{breg08}. We also assume $T=10^4$ K 
and density $n=0.2$ cm$^{-3}$.
The cloud is assumed to be
initially cylindrically shaped with both diameter and height equal
to 800 pc.  Figure \ref{fig:rainc} shows a sequel of snapshots taken
at different times and illustrating the dynamical interaction of the
falling cloud with the rising gas of the MGF.

\begin{figure}
\begin{center}
\psfig{figure=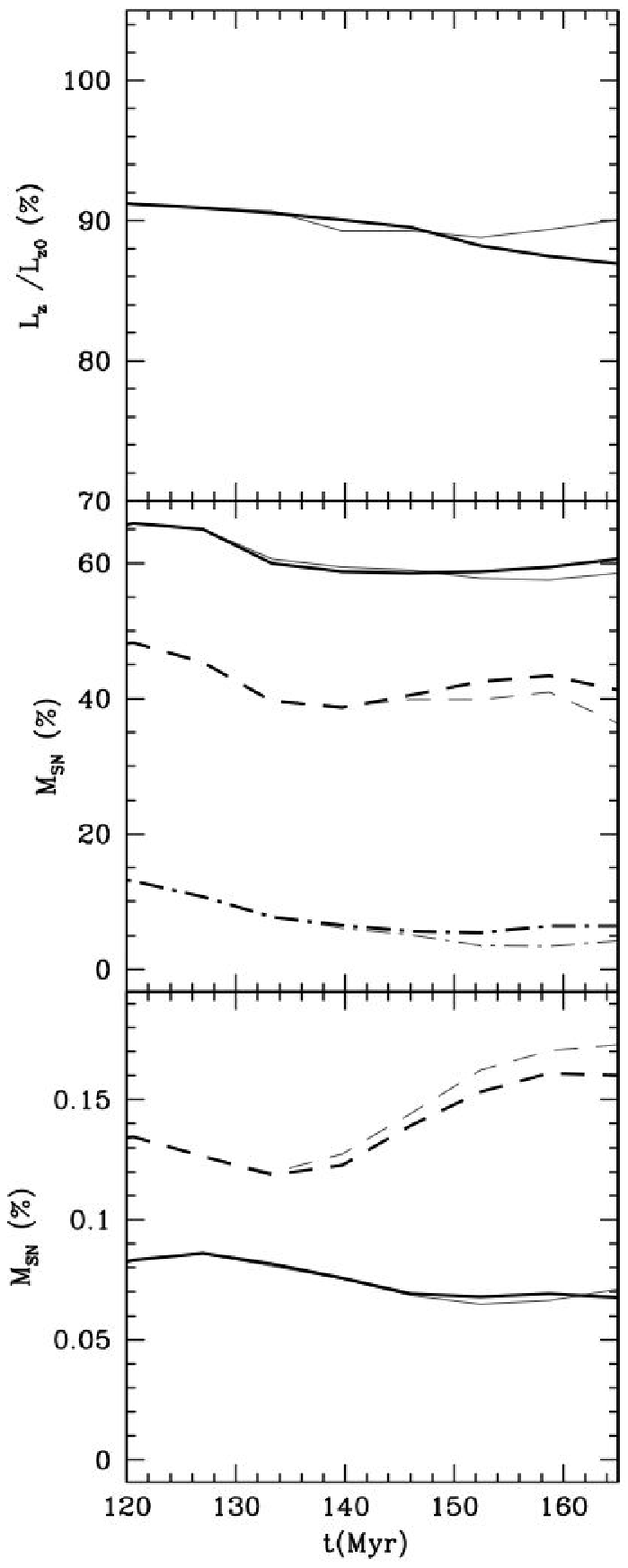,width=0.5\textwidth}
\end{center}
\caption{The same as in  Fig. \ref{fig:equil}, but for the cloud
model (thin lines) compared with the reference model (thick lines).
} \label{fig:equilc}
\end{figure}

The figure shows that the cloud actually interacts only with a
fraction of all the MGF gas. In fact, the cloud diameter is $\sim
1.2$ kpc (the cloud expands somewhat during its trajectory before
entering in equilibrium with the ambient gas), and its covering
factor relative to the active area is $\sim 0.15$. As a consequence,
the general characteristics of the fountains are not greatly altered
by the cloud interaction, as clearly illustrated in Fig.
\ref{fig:equilc}.

\section{Conclusions}
\label{sec:conc}

In Paper I we studied the evolution of a single fountain,
i.e. the flow generated by a single SN II association, in order to
understand the influence of a number of factors (such as the Galactic
rotation, the distance from the centre of the Galaxy, and the presence
of a hot Galactic halo) on the fountain gas circulation and on the
properties of the clouds forming in this process. In the present paper,
we have instead considered multiple GFs, i.e., the action (and interaction) of
several GFs powered by SN II associations sufficiently close one to
another in space and time. As in Paper I, we focused on a model
located at $R=8.4$ kpc, the solar Galactocentric distance. Our
findings can be summarized as follows:
\begin{enumerate}
\item The spatial and temporal random sequel of SNe II produces holes
  on the disk, giving ``punctured'' look to it, qualitatively similar
  to what is
  observed in many galaxies, as NGC 2403 \citep{frat02}. The detailed ``topology''
of the gaseous disk depends on several factors such as the SN II rate,
the size of
  the holes (which in turn depends on the clustering of the SNe
  and the local density of the ISM -- See Section 4),
  and the speed at which they disappear once the SNe
  responsible of their formation stop exploding.
\item As expected, a larger amount of extraplanar gas can be found at larger
  heights relative to a SGF model. However, and somewhat more unexpectedly, 
  the fountain flow remains
  radially localized, just as in the SGF case. The radial dispersion of the MGF
  gas, instead, is smaller than in the case of a SGF. 
  Given the absence of significant
  radial flows, the fountains play a negligible role in shaping the
  radial profile of the disk chemical abundance, as required by
  chemical models which ignore the hydrodynamics of the ISM
  (e.g.,\citep{cemafr07}).
\item Multiple fountains can form extraplanar gas at larger heights
  compared to SGFs. However, despite their higher energy budget, the
  MGFs are not able to produce clouds at an altitude larger than $\sim
  3.5$ kpc above the Galactic plane. Importantly, these clouds are mainly made of
  disk ISM (thus sharing the solar metallicity), rather than cooled SN ejecta
or halo gas. They fall back toward
  the Galactic plane with velocities lower than 100 km $^{-1}$.  From
  the considerations above we conclude that the gas condensations
  rising in our simulations can be identified with IVCs (rather than
  HVCs). In this scenario, the IVCs have a local origin
  \citep[c.f.][]{wak08}.
\item After the break out the multiple fountains protrude into
  the hot gas surrounding the Galaxy and an energy exchange is
  established between them and the hot gas. The diffuse fountain
  gas which does not participate of the cloud formation stays over the
  disk and becomes part of the halo gas. Its thermal energy remains in the hot halo.
  On the other
  hand, part of this coronal gas cools radiatively and condenses on
  the clouds. We find that the final balance is in
  favor of the halo, which is heated by the SNII with an efficiency of $\sim 10$\%.
  While this moderate value implies that SN feedback is probably not able to solve
  the ``overcooling'' problem that arises in the cosmological
  context of galaxy formation \citep[e.g.][]{nav97, tor04}
  it is more that enough to mantain an extended hot gaseous halo and cold/warm
  extraplanar gas around typical spiral galaxies.

\item 
  Our hydrodynamical simulations indicate that the extraplanar gas generated
  by the fountains accretes
  mass from the infalling IGM at a rate similar to that inferred by \citet{frabi08}.
  Though our
  simulations are restricted to a small area of the Galactic disk and
  therefore, are not suitable to reproduce the rotational curve of the
  entire thick disk, they show that the presence of an
  external gas infall may help to slow down the rotation of the gas in
  clouds (Fig. 16) and thus the amount of angular momentum
  that is transferred to the halo. However, higher resolution simulations 
  are still needed
  in order to explore the role of the K-H instabilities upon the
  clouds fragmentation and disintegration which have been neglected by
  \citet{frabi08} and which are not properly resolved in our
  simulations.

Finally, we considered the accretion of single, isolated
clouds. Our simulations show that this kind of accretion does not
alter the general characteristics of the MGFs that arise from the
disk.

\end{enumerate}

\section*{Acknowledgments}
E.M.G.D.P. acknowledges the partial support from grants of the
Brazilian Agencies FAPESP and CNPq.



\bibliographystyle{mn2e} 
\bibliography{fontane_ref}

\label{lastpage}
\end{document}